\documentclass[sn-nature]{sn-jnl}

\usepackage{graphicx}%
\usepackage{multirow}%
\usepackage{amsmath,amssymb,amsfonts}%
\usepackage{amsthm}%
\usepackage{mathrsfs}%
\usepackage[title]{appendix}%
\usepackage{xcolor}%
\usepackage{textcomp}%
\usepackage{manyfoot}%
\usepackage{booktabs}%
\usepackage{algorithm}%
\usepackage{algorithmicx}%
\usepackage{algpseudocode}%
\usepackage{listings}%
\usepackage{hhline,multirow}

\begin{document}

\title[Article Title]{A binary system in the S~cluster close to the supermassive black hole Sagittarius A*}

\author*[1]{\fnm{Florian} \sur{Peißker}}\email{peissker@ph1.uni-koeln.de}

\author[2,1]{\fnm{Michal} \sur{Zaja\v{c}ek}}

\author[1]{\fnm{Lucas} \sur{Labadie}}

\author[1]{\fnm{Emma} \sur{Bordier}}

\author[1,3]{\fnm{Andreas} \sur{Eckart}}

\author[1]{\fnm{Maria} \sur{Melamed}}

\author[4]{\fnm{Vladim\'{\i}r} \sur{Karas}}

\affil*[1]{\orgdiv{1.Physikalisches Institut}, \orgname{Universität zu Köln}, \orgaddress{\street{Zülpicher Str. 77}, \city{Cologne}, \postcode{50937}, \country{Germany}}}

\affil[2]{\orgdiv{Department of Theoretical Physics and Astrophysics}, \orgname{Masaryk University}, \orgaddress{\street{Kotlářská 2}, \city{Brno}, \postcode{61137}, \country{Czech Republic}}}

\affil[3]{\orgdiv{Max-Plank-Institut für Radioastronomie}, \orgname{Max-Planck-Gesellschaft}, \orgaddress{\street{Auf dem Hügel 69}, \city{Bonn}, \postcode{53121}, \country{Germany}}}

\affil[4]{\orgdiv{Astronomical Institute}, \orgname{Czech Academy of Sciences}, \orgaddress{\street{Bo\v{c}n\'{\i} II 1401}, \city{Prague}, \postcode{141\,00}, \country{Czech Republic}}}

\abstract{High-velocity stars and peculiar G~objects orbit the central supermassive black hole (SMBH) Sagittarius~A* (Sgr~A*). {Together, the G~objects and high-velocity stars constitute the S cluster.} In contrast with theoretical predictions, no binary system near Sgr~A* has been identified. Here, we report the detection of a spectroscopic binary system {in the S~cluster} with the masses of the components of $2.80\,\pm\,0.50\,\mathrm{M_{\odot}}$ and $0.73\,\pm\,0.14\,\mathrm{M_{\odot}}$, {assuming an edge-on configuration}. Based on periodic changes in the radial velocity, we find an orbital period of 372\,$\pm$\,3 days for the two components. The binary system is stable against the disruption by Sgr~A* due to the semi-major axis of the secondary being 1.59$\,\pm\,$0.01~AU, which is well below its tidal disruption radius of approximately 42.4~AU. The system, known as D9, shows similarities to the G~objects. We estimate an age for D9 of $2.7^{+1.9}_{-0.3}\,\times\,10^6$~yr that is comparable to the timescale of the SMBH-induced von Zeipel-Lidov-Kozai cycle period of about $10^6$~yr, causing the system to merge in the near future. Consequently, the population of G~objects may consist of pre-merger binaries and post-merger products. {The detection of D9 implies that binary systems in the S~cluster have the potential to reside in the vicinity of the supermassive black hole Sgr~A* for approximately $10^6$~years.}}


\maketitle

\section*{}\label{sec1}
The central parsec around the supermassive black hole (SMBH) Sgr~A* contains a large number of stars that constitute the Nuclear Star Cluster (NSC) \cite{Schoedel2009}, which is one of the densest and most massive stellar systems in the Galaxy. These stars vary in terms of their ages, masses, sizes, and luminosities \cite{Schoedel2010}. In the vicinity of Sgr~A* {of about 40 mpc}, there is a high concentration of stars \cite{Ali2020} that orbit the black hole at velocities of up to several thousand km/s \cite{Eckart1996Natur, Ghez1998} inside the S~cluster. The presence of stars close the Sgr~A* is not surprising because it was expected that old and evolved stars would gradually descend towards Sgr~A* due to the cluster relaxation timescale of {about} $10^{10}$~yr \cite{Morris1993}. This is because star formation is significantly inhibited by tidal forces and high energetic radiation in the vicinity of the SMBH. In fact, \cite{Habibi2019} identified a cusp of late-type stars with stellar ages of $>3\,\times\,10^{9}$~yr. Interestingly, these late-type stars coexist with massive early-type S~cluster members that exhibit an average age of approximately $4-6\,\times\,10^{6}$~yr \cite{Lu2013, Habibi2017}, resulting in the formulation of the ``paradox of youth'' \cite{Ghez2003}. Until now, no companions have been identified for these young B-type stars \citep{Chu2023}, although binary rates close to 100$\%$ have been proposed \citep{Offner2023}.
Therefore, the presence of binary systems in the S~cluster is a crucial question to investigate the dynamical evolution of stars in the vicinity of Sgr~A* \citep{Michaely2023, Gautam2024}. Given that the evolution of high-mass stars is altered by their binary interactions \citep{Sana2012}, it is important to understand the prevalence of putative binary systems in this cluster.\newline
In this work, we present the detection of a spectroscopic binary in the S~cluster. Based on the photometric characteristics of the binary system, known as D9, it can be considered to be a member of the G~object population \citep{Ciurlo2020, Peissker2024}. The age of the system is about $2.7\,\times\,10^6$yr, which is comparable to the von Zeipel-Lidov-Kozai cycle period of approximately $10^6$ years. The dusty source D9 is most likely composed of a Herbig Ae/Be star associated with the primary. The lower-mass companion can be classified as a T-Tauri star. In the near future, the binary may undergo a merging event due to the ongoing three-body interaction of the system with Sgr~A*. The uncertain nature of the G~objects can thus be resolved, at least in part, thanks to the binary system D9 whose imminent fate appears to be a stellar merger.

\section*{Results}

\subsection*{Observations}
\begin{figure}[ht!]%
\centering
\includegraphics[width=1.\textwidth]{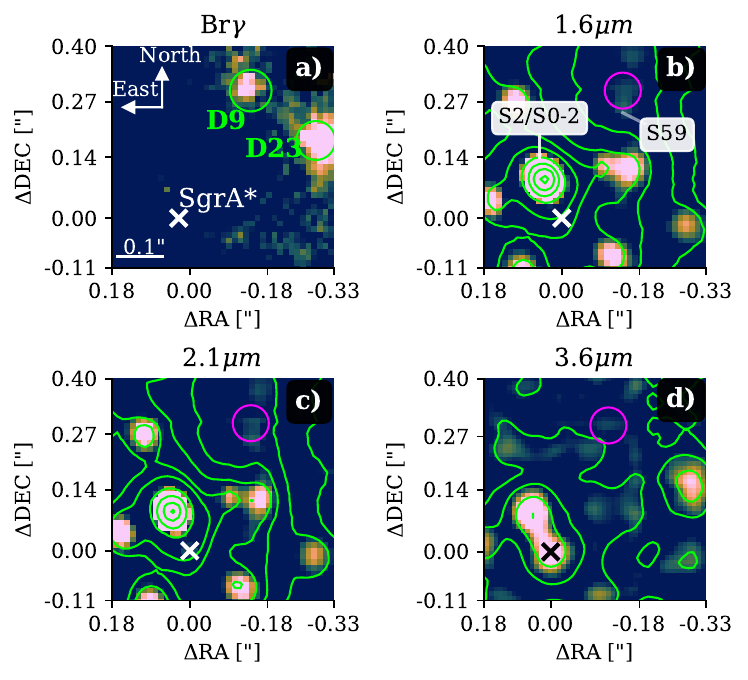}
\caption{{\bf Detection of the D9 system close to Sgr~A* in 2019}. Subplot (a) shows the Doppler-shifted Br$\gamma$ line map extracted from the H+K SINFONI data cube with a corresponding wavelength of 2.1646$\,\mathrm{\mu m}$ (vacuum wavelength 2.1661$\,\mathrm{\mu m}$). Subplot (b) and (c) shows the near-infrared H (1.6$\,\mathrm{\mu m}$) and K (2.1$\,\mathrm{\mu m}$) band data observed with SINFONI. Subplot (d) denotes the mid-infrared L (3.76$\,\mathrm{\mu m}$) band observation carried out with NIRC2. Sgr~A* is marked with a $\times$, D9 is encircled in every plot. Due to its main sequence character, the marked close-by star S59 can only be observed in the H and K bands. On the contrary, the brightest K band source of the S~cluster, S2/S0-2 can be observed in every shown infrared band. To increase contrast, an image sharpener is applied suppressing expansive {point spread function (PSF)} wings. To emphasize the astrometric robustness of the image sharpener, we adapt the lime-colored contour lines from the non-sharpened data. The contour line levels in panel b) are at $10\%$-$80\%$ of the peak intensity of S2, increasing in $5\%$ steps. In panel c), the contour lines are set at $20\%$-$100\%$ of the peak intensity of S2, separated by $10\%$. For panel d), the contour lines are set to $85\%$, $90\%$, $95\%$, and $100\%$ of the peak intensity of S2. The labels of the axis indicate the distance to Sgr~A* located at $\Delta$RA=$0.00"$ and $\Delta$DEC=$0.00"$. In any plot shown, north is up, and east is to the left.}
\label{fig:d9_multiwavelength_detection}
\end{figure}

Using archival data observed with the decommissioned near-infrared integral field unit (IFU) of {Spectrograph for INtegral Field Observations in the Near Infrared (SINFONI, mounted at the Very Large Telescope) \citep{Eisenhauer2003, Bonnet2004}} in the H+K band ($\rm 1.4-2.4\,\mathrm{\mu m}$) between 2005 and 2019, we investigate the blue-shifted Brackett$\gamma$ (Br$\gamma$) emission of the source D9 (Fig. \ref{fig:d9_multiwavelength_detection}), which is part of the G~object population in the S~cluster \cite{Peissker2020b, Ciurlo2020, Peissker2024}. In addition, we include recent Enhanced Resolution Imaging Spectrograph (ERIS) observations carried out by the ERIS Team as part of the commissioning run in 2022 \citep{Davies2023}. For the analysis of the three-dimensional data cubes that consist of two spatial and one spectral dimensions, we perform standard reduction steps (flat-fielding, dark, and distortion corrections). We obtain single barycentric and heliocentric corrected data cubes that are stacked for each year individually to construct a final mosaic of the entire S~cluster region.
Based on the best-fit Keplerian solution, we obtain an estimate of the periapse distance of the D9 system from Sgr~A* of 29.9~mpc (0.75~arcseconds) adopting M$_{\rm SgrA*}\,=\,4\,\times\,10^6\,\mathrm{M_{\odot}}$ and $8$~kpc for the mass and the distance of Sgr A*, respectively \cite{Peissker2022, eht2022}. Furthermore, we find a close to edge-on orbital inclination of $\rm (102.55\,\pm\,2.29)^{\circ}$. With an eccentricity of 0.32 and a semi-major axis of 44~mpc, D9 qualifies as an S~cluster member with orbital parameters comparable to other S~stars \cite{Gillessen2017, Ali2020}.
Due to the orbit of the B2V star S2 (S0-2) that intercepts with the trajectory of D9, we focus on the data set of 2019 to identify a continuum counterpart in the H and K band to the Br$\gamma$ line-emitting source.

\subsection*{Magnitudes}

To increase the photometric baseline, we incorporate {Near-infrared Camera 2 (NIRC2, mounted at the Keck telescope)} L band imaging data from 2019 to cover the near- and mid-infrared \citep{Matthews1994}. {The science-ready data was downloaded from the Keck Observatory Archive \citep{Tran2016}.} Due to the high stellar density of the S~cluster \cite{Eckart2013}, dominant {point spread function (PSF)} wings are a common obstacle that hinders confusion-free detection of fainter objects such as G1 \cite{Witzel2017}, DSO/G2 \cite{peissker2021c}, or D9 \cite{Peissker2020b}. Therefore, we used an image sharpener on the continuum data of 2019 to reduce the impact of the challenging crowding situation in the S~cluster (Supplementary Fig. 1 and Supplementary Table 1). With this procedure, we enhance fainter structures but preserve the photo- and astrometric aspects of the input data. To emphasize the robustness of the image sharpener, we invoke the contour lines of the input data as a comparison, as demonstrated in Fig. \ref{fig:d9_multiwavelength_detection}. Analyzing the displayed extinction corrected data (Supplementary Table 2), we find $\rm H-K=1.75\pm\,0.20$ and $\rm K-L=2.25 \pm 0.20$ colors for D9 suggesting photometric similarities with D2 and D23 \cite{Peissker2020b}. The latter two sources are believed to be associated with young T Tauri or low-mass stars \cite{Scoville2013, Valencia-S.2015, Ciurlo2020}. Due to these photometric consistencies (Supplementary Fig. 2), we tested the hypothesis using a Spectral Energy Distribution (SED) fitter. 

\subsection*{Spectral Energy Distribution}

\begin{figure}[ht!]%
\centering
\includegraphics[width=0.8\textwidth]{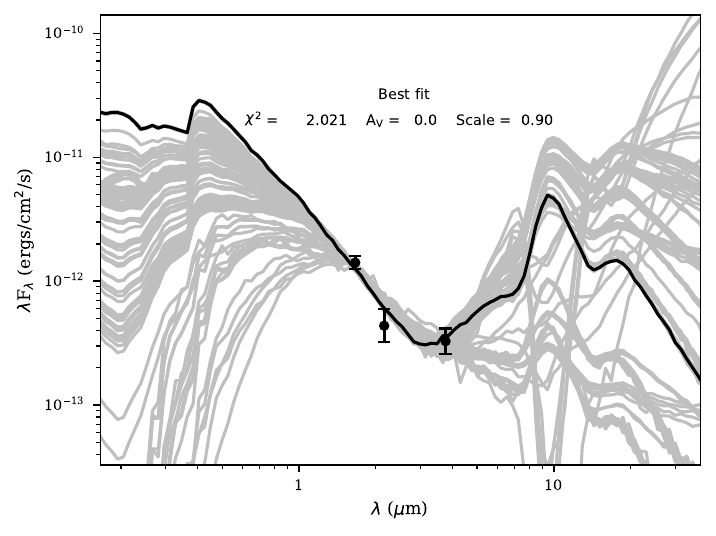}
\caption{{\bf Spectral Energy Distribution of the D9 system.} The extinction corrected data points refer to the flux density values in the H, K, and L band observed with SINFONI and NIRC2. We use 10$^4$ individual models to find the best fit of the data shown with grey lines. The final best-fit result is depicted with a black line. Based on the shown fit, the properties of the primary of the D9 binary system are derived and listed in Table \ref{tab:d9_parameter}. The uncertainties of the data points are estimated from the photometric variations along the source. Based on the reduced $\chi^2$ value of about 2, the displayed best-fit solution was selected.}
\label{fig:d9_sed}
\end{figure}

The SED fitter \cite{Robitaille2017} applies a convolving filter to the individual values to reflect on the response function of the instrument filter. Because the photometric system of SINFONI is based on the Two Micron All Sky Survey (2MASS) data base, we select the corresponding filters ``2H" and ``2K". For the NIRC2 MIR data, we use the United Kingdom Infrared Telescope (UKIRT) L' band filter because it is based on the Mauna Kea photometric system \cite{Leggett2003}. With these settings, the fitter compares models with the input flux (Fig. \ref{fig:d9_sed}) where we limit the possible output that satisfies $\Delta \chi^2\leq3$. These models represent young stellar objects (YSOs) and are composed of a stellar core, an accretion disk, and a dusty envelope. These typical components constitute a YSO and can be traced in the near- and mid-infrared parts of the spectrum. As input parameters, we used the H ($0.8\,\pm\,0.1$~mJy), K ($0.3\,\pm\,0.1$~mJy), and L ($0.4\,\pm\,0.1$~mJy) band flux densities estimated from the continuum detection presented in Fig. \ref{fig:d9_multiwavelength_detection}. 
Considering common YSO models, the H and K band emission traces the core components of the system, whereas the L band emission can be associated with a dusty envelope. Based on a photometric comparison with 10$^4$ individual models, the best-fit of the SED fitter results in {a stellar temperature of} $1.2\times10^4\,{\rm K}$ and a corresponding luminosity of approximately $ 93\,L_{\odot}$, which are associated with a stellar mass of $2.8\pm0.5\,\mathrm{M_{\odot}}$ (see Table \ref{tab:d9_parameter}). 

\subsection*{Periodic pattern}

\begin{figure*}[ht!]%
\centering
\includegraphics[width=1.\textwidth]{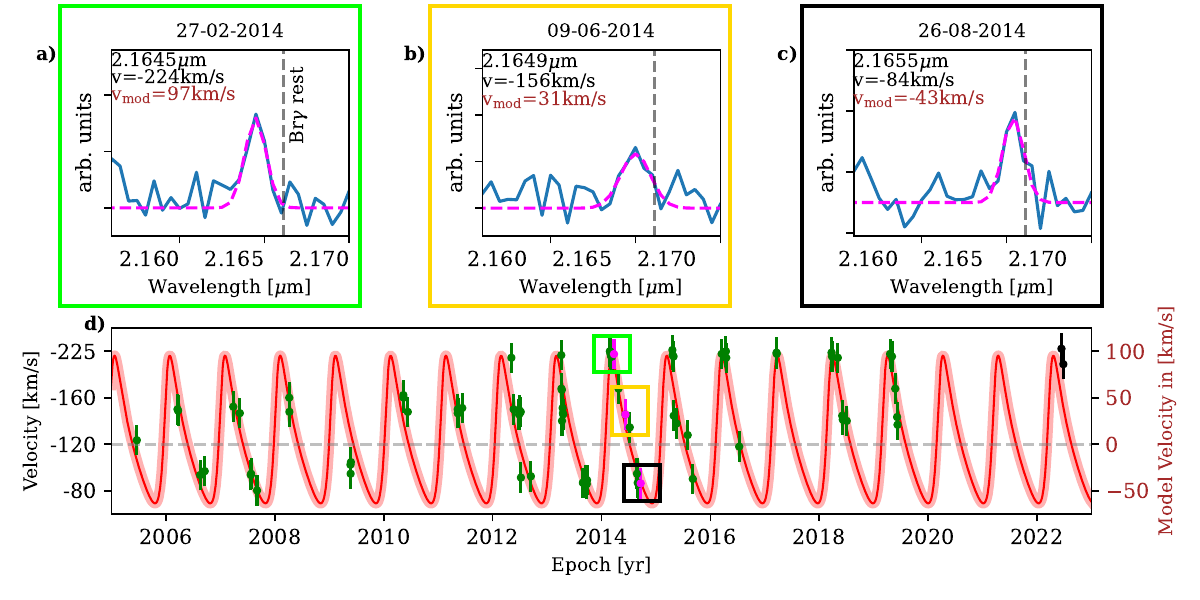} 
\caption{{\bf Radial velocity of D9 between 2005 and 2022 observed with SINFONI and ERIS.} In subplots (a), (b), and (c), we display three selected nights to show the variable Br$\gamma$ emission line with respect to the rest wavelength at 2.1661~$\mathrm{\mu m}$. The top three plots correspond to the same colored boxes as in the radial velocity evolution model shown in subplot (d). We have indicated the exact data point using magenta color. In subplot (d), the SINFONI data is indicated in green, the two ERIS observations from 2022 are highlighted in black. Due to the decommission, no high-resolution spectroscopic data are available between 2020 and 2021. In addition, the usual observation time for the Galactic center at Cerro Paranal (Chile) is between March and September, which explains the limited phase coverage. All data points in the radial velocity subplot (d) correspond to a single night of observation. The velocities in the left y-axis are related to the observed blue-shifted Br$\gamma$ emission lines. Due to data processing, these values are shifted and arranged to an estimated zero-velocity baseline (see the right y-axis). The uncertainties of the individual data points are calculated from the root-mean-square (RMS) deviation (see Table \ref{tab:d9_parameter}).}
\label{fig:d9_rv}
\end{figure*}

While finalizing the analysis of D9, a pattern of radial velocity came to our attention. By inspecting the SINFONI mosaics that depict every observed night between 2005 and 2019, we found a clear periodic signal shown in Fig. \ref{fig:d9_rv} between $-80$~km/s and $-225$~km/s using the Doppler-shifted Br$\gamma$ emission line with respect to its rest wavelength at $2.1661\,\mathrm{\mu m}$. A comparison of the periodic pattern of D9 with the Doppler-shifted Br$\gamma$ emission line of D23 demonstrates that the signal is not an artefact (Supplementary Fig. 3).
From the orbital fit and the related inclination of $\rm i =(102.55\pm2.29)^{\circ}$, we know that D9 is moving on an almost edge-on orbit with a proper motion of v$_{\rm prop}\,=\,249.43\,\pm\,5.01$~km/s. Since S2 (S0-2) moves with a proper motion of almost 800~km/s \cite{Eckart2002}, the comparable slow velocity of D9 implies that the intrinsic RV baseline v$_{\rm base}$ of the system, estimated with (v$_{\rm min}+$v$_{\rm max}$)/2, will not change significantly between 2005 and 2019. We normalize all observed velocities v$_{\rm obs}$ to this baseline with v$_{\rm obs}$-v$_{\rm base}$ to obtain v$_{\rm norm}$, which is the input quantity for the fit of the binary system performed with Exo-Stricker \cite{exostriker2019}. Due to the poor phase coverage before 2013, we split the data to perform an independent sanity check. The fit displayed in Figure \ref{fig:d9_rv} resembles the epochs between 2013 and 2019, where we used a false-alarm probability of $10^{-3}$ similar to that used by \cite{Gautam2024}. The data baseline between 2005 and 2012 represents a non-correlated parameter to the Keplerian model of the binary provided by Exo-Striker, which is in agreement with the fit that is based on the epochs between 2013 and 2019 (Fig. \ref{fig:d9_rv}). With a similar motivation, we incorporate the ERIS observations from 2022 that show a satisfactory agreement with the RV model and the expected LOS velocity of the binary, consisting of a primary and a secondary.
Regarding the possible impact of a variable baseline v$_{\rm base}$ (i.e., the LOS velocity v$_{\rm obs}$ of D9 increases), we measure a difference of $\rm \pm\,15$~km/s between 2013 and 2019, which is consistent with the estimated uncertainty of $\rm \pm\,17$~km/s from the fit. We conclude that a variation of v$_{\rm base}$ over the complete data baseline is inside the uncertainties and does not impact the analysis significantly. However, a forthcoming analysis of the binary system D9 should take this adaptation into account because it is expected that an alteration of the intrinsic LOS velocity will exceed the uncertainty range of the individual measurements of the periodic signal within the next decade.\newline
{In the subsequent analysis, we will refer to the primary as D9a, whereas the secondary companion will be denoted as D9b. With the binary orbiting Sgr~A*, this three-body system is divided into an inner and outer binary. The inner binary describes D9a and D9b, while the outer one represents the D9 system orbiting Sgr~A*.}\newline
The best-fit result includes an offset of v$_{\rm base}$ with RV$_{\rm off}$ = -29.19 $\pm$ 3.00~km~s$^{-1}$ due to the eccentricity of the secondary e$_{\rm D9b}$ of 0.45$\,\pm\,$0.01, which causes an asymmetric distribution of the LOS velocity around the baseline. With this offset, we obtain v$_{\rm mod}$ = v$_{\rm norm}$ + RV$_{\rm off}$ as displayed in Fig. \ref{fig:d9_rv}.
The related Keplerian parameters of the secondary orbiting its primary are listed in Table \ref{tab:d9_parameter}. 
From the fit but also evident in the periodic RV data points (Fig. \ref{fig:d9_rv}), we find an orbital period for the secondary of {P$_{\rm D9b}$ = 372.30$\,\pm\,$3.65~days = 1.02$\pm$0.01~yr, which can be transferred to a total mass of the system of about M$_{\rm bin}$=$3.86\,\pm\,0.07\,\mathrm{M_{\odot}}$, considerably above the derived D9 (i.e., the primary) mass of M$_{\rm D9a}\,$=$\,2.8\,\pm\,0.5\,\mathrm{M_{\odot}}$}. The difference in mass for M$_{\rm D9a}$ and M$_{\rm bin}$ cannot be explained solely by the uncertainty range. However, inspecting $\rm m \sin(\mathrm{i}_{\rm D9b})\,=\, 0.73\,\mathrm{M_{\odot}}$ and the assumed inclination of the secondary of 90$^{\circ}$ results in the maximum mass of the companion. The assumed inclination of the secondary is motivated by an almost edge-on orbit of D9 (Table \ref{tab:d9_parameter}).
Although the circumprimary disk does not necessarily have to be aligned with the orbit of the binary as is found for T-Tauri systems \cite{Bally2005}, surveys of Herbig Ae/Be stars suggest a tendency towards a coplanar arrangement \cite{Baines2006}. Assuming that the orbit of the secondary is approximately aligned with the circumprimary disk with an intrinsic inclination of the primary D9a of i$_{\rm intrinsic}\,=\,(75\pm19)^{\circ}$ (Table \ref{tab:d9_parameter}), we are allowed to transfer the related uncertainties to $\rm m\sin(\mathrm{i}_{\rm D9b})$. Following this assumption, we find a mass for the secondary of $\rm M_{\rm D9b}\,=\,0.73\,\pm\,0.14\,\mathrm{M_{\odot}}$ consistent with the derived primary mass of $\rm M_{\rm D9a}\,=\,2.8\,\pm\,0.5\,\mathrm{M_{\odot}}$ and the total mass $\rm M_{\rm bin}\,$=$\,3.86\,\pm\,0.07\,\mathrm{M_{\odot}}$ of the system.

\section*{Discussion}

\subsection*{Radiation mechanism}

Taking into account the periodic variation of Br$\gamma$ emission, we want to highlight three different scenarios as a possible origin of the periodic Br$\gamma$ signal.\newline
Firstly, the emission of the Br$\gamma$ line is solely the result of a combination between the gaseous accretion disk and stellar winds of the primary \cite{Tanaka2016, Tambovtseva2016}. In this scenario, the secondary disturbs this emission by its intrinsic Keplerian orbit around the primary.\newline
Secondly, a possible origin of the Br$\gamma$ line could be the presence of a circumbinary disk around the D9 binary system enveloping the primary and the secondary. In this case, the interaction between the primary with the secondary allows inward gas streams from the circumbinary disk resulting in the observed periodic Br$\gamma$ line \cite{Fateeva2011}.\newline 
The third and foremost plausible scenario is the interaction between two accreting stellar objects. It is well known that especially Herbig Ae and T-Tauri stars exhibit prominent Br$\gamma$ emission lines associated with accretion mechanisms \cite{Muzerolle1998, Grant2022}. For instance, a radial shift of the accretion tracer has been observed for the DQ Tau binary system \cite{Fiorellino2022}. It has been proposed that this resonance-intercombination may be explained by stellar winds of the secondary \cite{Friedjung2010}. Due to Keplerian shear, line photons can escape the optically thick material and produce the RV pattern, as observed for the D9 binary system \cite{Horne1986}.

\subsection*{Stellar types of the primary and secondary}

Considering the presence of a primary and its companion, it is suggested that stellar winds interact with the Br$\gamma$ emission of the accretion disk(s) of the binary system \citep{Kraus2012,Tanaka2016} that gets periodically disturbed by the presence of the secondary \citep{Kraus2009,Frost2022}. Alternatively, the Br$\gamma$ emission line is produced by both the primary and secondary as it is observed for the Herbig Ae star HD 104237 with its T Tauri companion \citep{Garcia2013}. Comparing M$_{\rm D9a}$ with the total mass of M$_{\rm bin}$=$3.86\,\mathrm{M_{\odot}}$ of the system suggests that the secondary does not contribute significantly to the photometric measurements analyzed in this work. If it were not the case, the estimated mass for {the primary of the D9 system} of M$_{\rm D9a}$=$2.8\pm 0.5\,\mathrm{M_{\odot}}$ would be lower, while M$_{\rm D9b}=0.73\pm 0.14\,\mathrm{M_{\odot}}$ should be increased. Considering the estimated mass of the primary M$_{\rm D9a}$ and the fixed upper limit of M$_{\rm bin}$ based on the observed period, the secondary can be classified as a faint low-mass companion, suggesting a classification as a T-Tauri star \cite{Hartmann1998}. Considering the stellar mass, radius, and luminosity of the primary (Table \ref{tab:d9_parameter}), the system may be comparable to the young Herbig Ae/Be star BF Orionis, which is speculated to also have a low-mass companion \cite{Grinin2010}. On the basis of observational surveys, it is intriguing to note that most Herbig Ae/Be stars exhibit an increased multiplicity rate of up to 80$\%$ \cite{Baines2006, Wheelwright2010}. 
Another result of the radiative transfer model is the relatively small disk mass M$_{\rm Disk}$ of (1.61$\,\pm\,$0.02)$\,\times\,$10$^{-6}$~M$_{\odot}$, which could be interpreted as an indicator of the interaction between D9a and its low-mass companion D9b. Possibly, this ongoing interaction, but most likely the stellar winds of the S~stars \citep{Lutzgendorf2016}, will disperse the disk of D9a in the future \citep{Hollenbach1994, Johnstone1998, Messina2019, Alves2019}. {Using the derived luminosity and stellar temperature of D9a together with the evolutionary tracks implemented in PARSEC \citep{Bressan2012}, we estimated the age of the system of $2.7^{+1.9}_{-0.3}\,\times\,10^6$~yr.}


\subsection*{Migration scenario}

A potential migration scenario has been proposed by \cite{Fragione2018} and can be described as the triple-system hypothesis. In this scenario, a triplet system migrates towards Sgr~A* \cite{Bonnell2008, Jalali2014, Zajacek2017}, where the two companions are captured to form a binary. It is possible that the third companion may be ejected from the cluster and subsequently become a hyper-velocity (HV) star, as postulated by \cite{Hills1988, Yu2003}. A consequence of the disruption of the initial triplet is the resulting high eccentricity of the captured binary system close to unity \citep{Zajacek2014}. Since the derived {outer} eccentricity of the D9 system is $\rm e_{\rm D9a}\,=\,0.32\,\pm\,0.01$ (Table \ref{tab:d9_parameter}), we consider a migration channel different from the triple-system hypothesis. As proposed by \cite{Bonnell2008} and \cite{Jalali2014}, molecular clouds can migrate towards the inner parsec and consequently close to Sgr~A*. Speculatively, the D9 system could have formed during such an inspiral event. An additional implication based on the age estimate is the presumably evaporated circumbinary disk that enveloped the primary and secondary. The authors of \cite{Armitage2003} found that the timescales for dismantling the circumbinary disk scale with the separation between the primary and secondary. The relation can be formulated with t$_{\rm dis. time}\,\leq\,10^6$~yr$\,<\,$t$_{\rm D9a, age}\,=\,2.7^{+1.9}_{-0.3}\,\times\,10^6$~yr. The former relation is strengthened by the analysis of \cite{Alexander2012} who found that photoevaporative winds decrease the lifetime of the circumbinary disk as a function of distance. Independent of the stellar wind model, the author of \cite{Alexander2012} found that circumbinary disks evaporate between approximately $1-10\,\times\,10^6$~yr providing an explanation for the low disk mass of (1.61$\,\pm\,$0.02)$\,\times\,$10$^{-6}$~M$_{\odot}$ found for the D9 binary system. Between 2005 and 2022, the D9 binary system has remained stable in the gravitational potential dominated by Sgr~A*. This is evident from the observable periodic RV signal for almost 20 years. The conditions for the dynamical stability of the binary can be extracted directly from the Keplerian orbital fit and binary mass estimate by calculating the tidal (Hill) radius.\newline
For the periapse distance $r_{\rm p}$ of approximately 30~mpc corresponding to 6200~AU, we find the tidal (Hill) radius for D9 of $r_{\rm Hill}=r_{\rm p}(M_{\rm bin}/3M_{\rm SgrA*})^{1/3}\,=\,42.4$~AU. The effective orbital radius of the {inner} binary system is r$_{\rm eff}\,=\,1.26\,\pm\,0.01$~AU using the Keplerian orbital parameters {for the secondary} listed in Table \ref{tab:d9_parameter}. Therefore, the system remains in a stable, mildly eccentric orbit around Sgr~A*, and it can be further described as a hard binary. {This is expected since the evolution of the outer orbit of the system D9-Sgr~A* is dominated by the gravitational potential of the SMBH.} However, because of its age and potential interaction with the dense environment, the question of binary destruction timescales should be addressed. {It is plausible that the inner system D9a-D9b will actually become even harder and the components will eventually merge \citep{Rose2020}. This is due to the interaction of the D9 system with Sgr~A*, which acts as a distant massive perturber that alters the orbital parameters through the von Zeipel-Lidov-Kozai (vZLK) mechanism \cite{Zeipel1910, Lidov1962, Kozai1962}. Due to the young age of the binary system and, therefore, the short time in the S cluster (compared to the evolved stars), we will focus in the following section on the vZLK and other effects induced by the dark cusp of the S~cluster.}

\subsection*{Dynamical processes and stellar populations}

The lifetime of the D9 system with its estimated age of $2.7^{+1.9}_{-0.3}\,\times\,10^6$~yr and the semi-major axis of about $44\,{\rm mpc}$ can be compared with basic dynamical processes and their timescales as well as with other known stellar populations in the central parsec in the distance-timescale plot. Such a plot (see e.g. \cite{2006ApJ...645.1152H}) can be used to infer which dynamical processes can be relevant for the current and the future orbital evolution of D9 at a given distance. We use the timescales for the two-body non-resonant relaxation $\tau_{\rm NR}$, scalar and vector resonant relaxation $\tau_{\rm RR}^{\rm s}$ and $\tau_{\rm RR}^{\rm v}$, respectively, and the vZLK mechanism driving inclination-eccentricity oscillations taking place on the vZLK timescale $\tau_{\rm vZLK}$.
\begin{figure}[ht!]
    \centering
    \includegraphics[width=1.\textwidth]{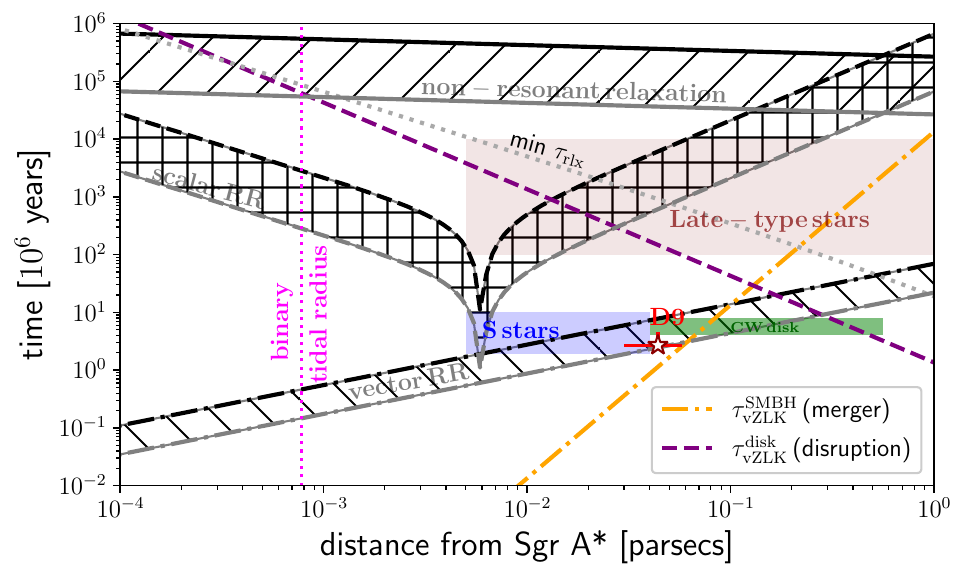}
    \caption{{\bf Distance and age of D9 in the context of basic dynamical processes and stellar populations in the Galactic center.} In terms of the semi-major axis, D9 is positioned in the outer part of the S cluster, close to the innermost part of the clockwise (CW) disk of OB/Wolf-Rayet stars. With its estimated age of $2.7^{+1.9}_{-0.3}\,\times\,10^6$~yr, its orbit around Sgr~A* can just be under the influence of the fast vector resonant relaxation (RR; shaded area stands for the vector resonant relaxation of a 1$\,M_{\odot}$ star and a $10\,M_{\odot}$ star represented by the top and the bottom lines, respectively). However, the scalar resonant relaxation (RR) and the non-coherent two-body relaxation have not had sufficient time to affect significantly the angular momentum and the orbital energy of the D9 system yet. Hence, D9 as a binary system is currently stable against the tidal disruption by Sgr~A* (vertical dotted magenta line denotes the binary tidal radius). A similar conclusion can be drawn with regard to the minimum relaxation time $\rm min\,\tau_{rlx}$ resulting from the dark cusp (illustrated by the orange dotted line). In addition, the von Zeipel-Lidov-Kozai (vZLK) mechanism that involves the SMBH-D9-CW disk ($\tau_{vZLK}^{\rm disk}$; dashed purple line) operates on a long timescale to cause the tidal disruption of the binary. On the other hand, in the hierarchical setup where the inner D9 binary orbits the SMBH, the corresponding vZLK timescale is comparable to the age of D9, which implies a likely merger (orange dash-dotted line).}
    \label{fig_timescales}
\end{figure}
In Fig.~\ref{fig_timescales}, we show the D9 system (red star), the timescales related to the dynamical processes, and the relevant stellar populations identified in the inner parsec: S cluster, clockwise (CW) disk, and late-type stars. For most of the timescales (non-resonant, scalar, and vector resonant relaxations), we need an estimate for the number of stars inside the given distance $r$ from Sgr~A*, $N(<r)$. For this purpose, we use the power-law mass density profile $\rho(r)=1.35\times 10^5 (r/2\,{\rm pc})^{-1.4}\,M_{\odot}\,{\rm pc^{-3}}$, whose power-law index is adopted from \cite{2018A&A...609A..26G} and the normalization coefficient is determined so that $M(<2\,{\rm pc})=2M_{\rm SgrA*}$, i.e. twice the Sgr~A* mass at the influence radius. We see that for the inferred age of D9, none of the relaxation processes is fast enough to change significantly the angular momentum magnitude, i.e. the eccentricity. Hence, the D9 binary is stable against disruption by Sgr~A* at the corresponding tidal radius $r_{\rm t}$ of about $161 (a_{\rm D9b}/1.59\,{\rm AU}) (M_{\rm SgrA*}/4\times 10^6\,M_{\odot})^{1/3} (M_{\rm bin}/3.86\,M_{\odot})^{-1/3}\,{\rm AU}\simeq 0.78\,{\rm mpc}$, for which the orbital eccentricity of $e\simeq 1-r_{\rm t}/a= 0.98$ would be required. Apart from nonresonant and scalar resonant relaxation processes, such a high eccentricity of the D9 orbit around Sgr~A* cannot be reached via the vZLK oscillations, where we consider Sgr~A* -- D9 as an inner binary and the CW disk as an outer perturber with the mass of $M_{\rm disk}\lesssim 10^4\,M_{\odot}$. With the mean distance {of the disk} $r_{\rm disk}$ of about $0.274\,{\rm pc}$ from D9, the corresponding vZLK cycle timescale is given by,
\begin{align}
    \tau_{\rm vZLK}^{\rm disk}&=2\pi \left(\frac{M_{\rm SgrA*}}{M_{\rm disk}} \right) \left(\frac{r_{\rm disk}}{a_{\rm D9a}} \right)^3 P_{\rm D9a}\,\notag\\
    &=2.6\times 10^8 \left(\frac{M_{\rm SgrA*}}{4\times 10^6\,M_{\odot}} \right) \left(\frac{M_{\rm disk}}{10^4\,M_{\odot}} \right)^{-1} \left( \frac{r_{\rm disk}}{0.274\,{\rm pc}}\right)^3 \times\,\notag\\
    &\times \left( \frac{a_{\rm D9a}}{0.044\,{\rm pc}}\right)^{-3} \left(\frac{P_{\rm D9a}}{432.35\,{\rm years}} \right)\,\text{yr}\,,
    \label{eq_vZLK_disk}
\end{align}
which is two orders of magnitude longer than the lifetime of D9 (see also Fig.~\ref{fig_timescales} for the radial dependency of $\tau_{\rm vZLK}^{\rm disk}$). {In Eq.~\eqref{eq_vZLK_disk}, we adopted the notation of the D9 orbital parameters as summarized in Table~\ref{tab:d9_parameter}.}\newline
When we concentrate instead on the other hierarchical three-body system -- the inner D9 binary and the outer binary D9--Sgr~A*, the inner binary components undergo the vZLK inclination--eccentricity cycles. The corresponding vZLK timescale then is,
\begin{align}
    \tau_{\rm vZLK}^{\rm SMBH}&=2\pi \left(\frac{M_{\rm bin}}{M_{\rm SgrA*}} \right) \left(\frac{a_{\rm D9a}}{a_{\rm D9b}} \right)^3 P_{\rm D9b}\,\notag\\
    &=1.1\times 10^6 \left(\frac{M_{\rm bin}}{3.86\,M_{\odot}} \right) \left(\frac{M_{\rm SgrA*}}{4\times 10^6\,M_{\odot}} \right)^{-1} \left( \frac{a_{\rm D9a}}{0.044\,{\rm pc}}\right)^3 \times\,\notag\\
    &\times \left( \frac{a_{\rm D9b}}{1.59\,{\rm AU}}\right)^{-3} \left(\frac{P_{\rm D9b}}{1.02\,{\rm years}} \right)\,\text{yr}\,,
    \label{eq_vZLK_SMBH}
\end{align}
which is within the uncertainties comparable to the age of D9. {In Eq.~\eqref{eq_vZLK_SMBH}, we adopted the notation of the parameters of both the D9 orbit around Sgr~A* and the binary orbit as summarized in Table~\ref{tab:d9_parameter}.} Hence, the system appears be detected  in the pre-merger stage. As the eccentricity of the D9 binary will increase during one vZLK timescale, the strong tidal interaction between the components during each periastron will perturb the stellar envelopes significantly, which will plausibly lead to the merger of both components once they are significantly tidally deformed \cite{Stephan2016}. Such a merger process is first accompanied by the Roche-lobe overflow of the stellar material from one of the components and then a subsequent merger of the stellar cores (see e.g. \cite{2003ApJ...582L.105S}). At the same time, the common envelope is progressively inflated to several thousand Solar radii. As it cools down, the infrared excess increases considerably. In this way, some or all of the G objects observed in the Galactic center could be produced and the D9 system would represent a unique pre-merger stage, which is also hinted by the smaller near-infrared excess in comparison with other G objects \cite{Peissker2024}.

\subsection*{Fate of the binary}

Due to the young age of the binary system and therefore its short time in the S cluster (compared to the evolved stars), we will first focus on the effect of the dark cusp. Old and faint stars have migrated into the S~cluster from a distance of a few parsecs \cite{Morris1993} and might alter the orbits of the young and bright cluster members \cite{Alexander2014, Habibi2019, Rose2020}. With the detection of the binary system D9, we convert its stellar parameters (Table \ref{tab:d9_parameter}) and age of $\rm T_{\rm D9a}\,=\,2.7^{+1.9}_{-0.3}\,\times\,10^6$~yr to a lower limit for the minimum two-body relaxation timescale of $\rm min\,t_{\rm rlx}\,=\,4.8(M_{\rm Sgr~A*}/M_{\rm bin})(a_{\rm D9b}/a_{\rm D9a})T_{\rm D9a}$ resulting in about $\rm 874\,\times\,T_{\rm D9a}$~yr \citep{Alexander2014}, equivalent to approximately $10^9$~yr exceeding the lifetime of the binary by three orders of magnitude. This suggests that the dark cusp does not have any significant imprint on the D9 system independent of its time in the cluster. Given that the assumed inclination is a geometrical parameter contingent upon the observer, it is reasonable to conclude that it will have, such as the dark cusp, no impact on the dynamical evolution of the binary system. We will now examine the evolutionary path that is described by the vZLK mechanism where D9 is the inner binary and D9-Sgr~A* represents the outer binary \citep{Stephan2016}. For this hierarchical setup, the vZLK timescale is $\tau_{\rm vZLK}^{\rm SMBH}\,=\,1.1\,\times\,10^6$~yr, see Fig. \ref{fig_timescales} and Eq.~\eqref{eq_vZLK_SMBH}, which is comparable with the approximate lifetime of the binary of $\rm T_{\rm D9a}\,=\,2.7\,\times\,10^6$~yr. It is reasonable to assume that the ongoing interaction between the primary, secondary, and Sgr~A* is reflected in altering the eccentricity of the D9 binary, which very likely results in a merger. This supports the idea that the G-object population \citep{Peissker2024} has a contribution from recently merged binary systems, as proposed by \cite{Ciurlo2020}. Considering the vZLK timescale $\tau_{\rm vZLK}^{\rm SMBH}$ of about $10^6$~yr and the age of D9 of $\rm 2.7^{+1.9}_{-0.3}\,\times\,10^6$~yr, the system {could have} migrated to its current location and may soon merge to become a G-object. D9 thus offers a glimpse on one potential evolutionary path of the S~stars. Taking into account that the bright and massive B-type S~stars with an average age of 6 $\times 10^6$yr \citep{Lu2013, Habibi2017} may have formed as binary systems \cite{Offner2023}, it is suggested that these young S~cluster members might have lost their putative companions in the immediate vicinity of Sgr~A* assuming an ex-situ formation. In \cite{Chu2023} and \cite{Gautam2024}, the authors explored the probability density for the young stars in and outside the S cluster. The authors propose that the probability of a binary system is significantly higher outside the central arcsecond ($\geq 72\,\%$ compared to $\leq 17\,\%$ at $68\%$ confidence interval). If we consider the recent detection of the new G~object X7.2 \citep{Ciurlo2023}, we estimate with $\rm R\,=\,N_{\rm B}/(2N_{\rm m})$ a binary fraction of the central 0.1 pc to be approximately $10\,\%$ using the Ansatz of \cite{Ciurlo2020}, where N$_{\rm B}\,=\,86$ \citep{Peissker2020b, Ciurlo2023} represents the assumed number of binaries and N$_{\rm m}\,=\,478$ \citep{Ciurlo2020} the amount of low-mass stars in the S cluster using the initial mass function derived by \cite{Lu2013}. This implies that the majority of expected binaries in the S cluster should be among the G~objects \cite{Peissker2020b}.\newline
{Regardless of the formation or migration scenarios, we can estimate that the B type stars of the S cluster reside in their environment for at least $\rm 1.1\,\times\,10^6$ years due to the absence of their expected companion stars \citep{Offner2023, Chu2023}. The} estimated vZLK timescale is compatible with the predicted decrease of binaries for a possible star-formation episode in the Galactic center 6$\,\times\,10^6$~yr ago \cite{Lu2013, Stephan2016}. This suggests that the vZLK mechanism may be the driving force of the decrease in binary fraction in the dense S~cluster \citep{stephan2019, Habibi2019, Rose2020}. 

\subsection*{Alternative explanations}

The number of detected binaries in the Galactic center is surprisingly low. Only five confirmed binaries have been found, which is, considering an approximate number of stars in the NSC of approximately $10^6$ \cite{Schoedel2009}, a negligible fraction of the overall population (Supplementary Table 3). Although the multiplicity fraction in the NSC should be higher \cite{Stephan2016, stephan2019}, other possible scenarios that explain the periodic RV pattern displayed in Fig. \ref{fig:d9_rv} should be taken into account.
One possible alternative explanation for the periodic variations of RV could be stellar pulsations \cite{Paxton2011}. This scenario was initially used to explain the photometric variability of IRS 16SW \cite{Ott1999, Depoy2004}. However, it was later confirmed that the Ofpe/WN9 star IRS 16SW is indeed a massive binary by conducting IFU observations with SINFONI \cite{Martins2006} analyzing the Br$\gamma$ emission line. Considering the binary period of the D9 system of about 372 days, stellar pulsations are rather unlikely, since they occur on daily timescales \cite{Paxton2019}. Alternatively, the Br$\gamma$ emission could be related to the rotation of the accretion disk of D9. Although ionized hydrogen and disk winds are associated with YSOs \cite{Kraus2012, Tanaka2016}, the dimensions of the disk itself and the spectral resolution of the instrument pose a strong constraint on the detectability of the system. 


\section*{Methods}\label{sec11}

\subsection*{Age of the system}

\begin{figure}[ht!]%
\centering
\includegraphics[width=1.\textwidth]{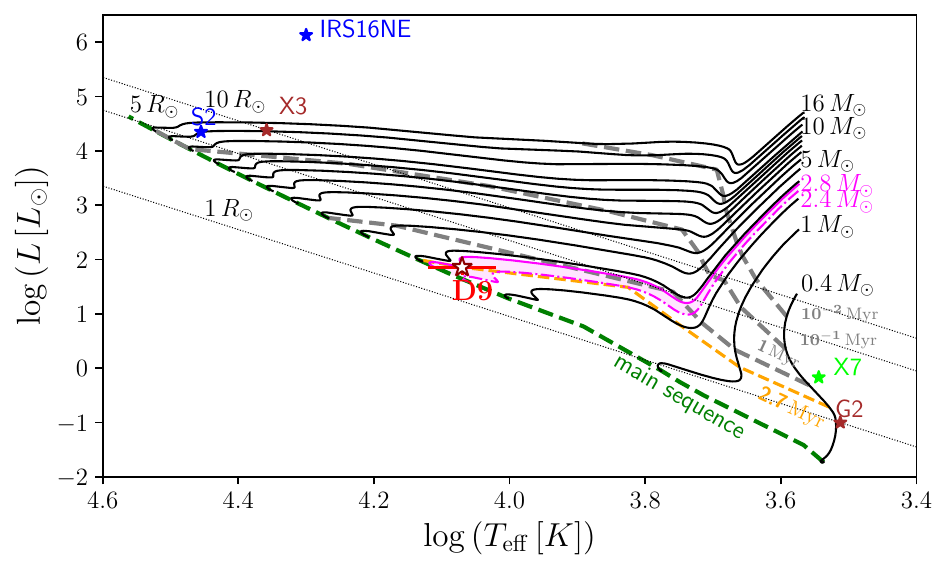}
\caption{{\bf Hertzsprung-Russel diagram using the evolutionary tracks based on the PARSEC stellar evolution model.} The D9 binary system is indicated by a {\bf red} star with the corresponding errorbars in the temperature-luminosity plot. The magenta-shaded area depicts the range of the masses of stars (2.4-2.8\,$M_{\odot}$), whose stellar evolution is consistent with the location of the D9 source at the time of $2.4-4.6\,\times\,10^6$~yr. The orange-dashed line represents the isochrone corresponding to 2.7 million years. For comparison, we implement known sources of the Galactic center, such as the putative high-mass YSO X3 \cite{peissker2023b}, the bow-shock source X7 \cite{peissker2021, Ciurlo2023}, dusty S cluster object G2 \cite{Shahzamanian2016}, and the massive early-type stars S2 \cite{2017ApJ...847..120H} and IRS16NE \cite{Najarro1997}.}
\label{fig:hr}
\end{figure}

For an age estimate of the D9 binary system, we use the temperature and radius listed in Table \ref{tab:d9_parameter} with stellar evolutionary tracks from PARSEC \cite{Bressan2012}. 
Considering the low mid-infrared flux in the L band of 0.4$\pm$0.1 mJy compared to the K band of 0.8$\pm$0.1 mJy, questions the proposed classification for D9 as a candidate Class I YSO as suggested by \cite{Peissker2024}. Taking into account the derived stellar mass of the system in combination with the hydrogen emission line, alternative explanations are required to classify the binary system. As outlined before, it is known that the Br$\gamma$ line is a tracer for accretion disks of Herbig Ae/Be stars \cite{Tambovtseva2016}. Similar to Herbig Ae/Be surveys \citep{Vioque2018}, we use the PARSEC isochrones \citep{Bressan2012} to estimate the age of D9 (Fig. \ref{fig:hr}).
We find an age of the D9 system of $2.7^{+1.9}_{-0.3}\,\times\,10^6$~yr (Fig. \ref{fig:hr}), which is, in combination with the high binary rate \cite{Baines2006, Wheelwright2010, Vioque2018}, typical for Herbig Ae/Be stars. 
This age estimate implies an ex-situ formation scenario because the dominant winds of the massive stars inside the S~cluster would have photoevaporated the required star formation material in the first place \citep{McKee2007, Lutzgendorf2016}. The stellar evolution model is in agreement with common stellar parameters of Herbig Ae/Be stars \citep{Vioque2018, Wichittanakom2020} that are derived from the Gaia Data Release 2 \citep{Gaia2016, Gaia2018}.

\subsection*{Keplerian orbit}

\begin{figure}[ht!]%
\centering
\includegraphics[width=.8\textwidth]{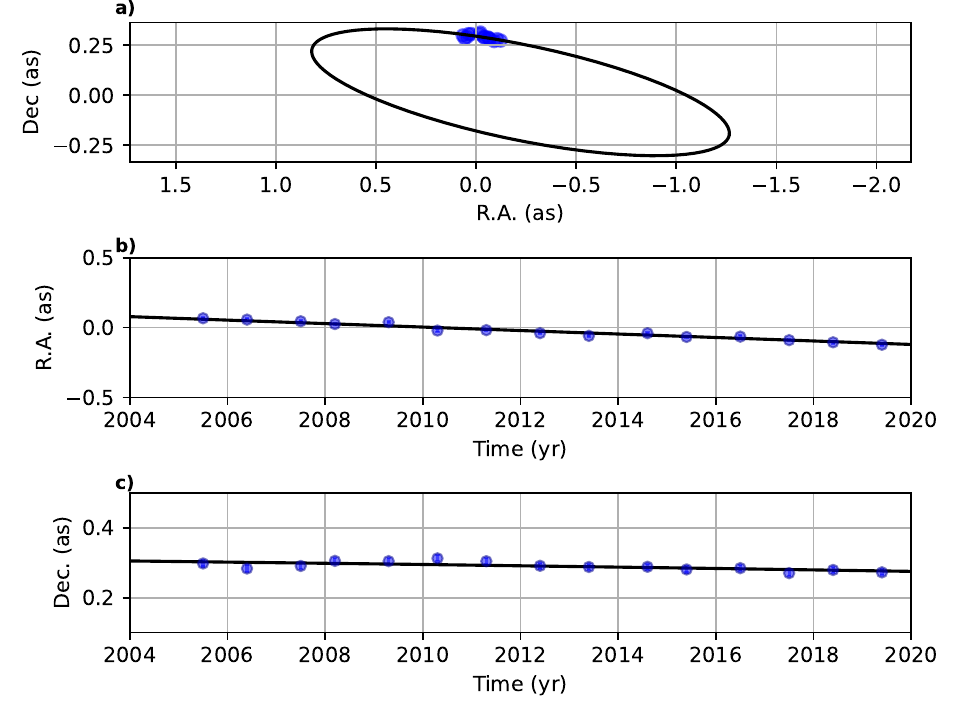}
\caption{{\bf Keplerian orbit of the D9 system.} In subplot (a), the projected on-sky trajectory of the D9 binary system is shown. Subplot (b) and (c) shows the R.A. and DEC. position as a function of time. In subplot (b) and (c), the low proper motion is eminent. Every blue-colored data point in this figure is related to one observational epoch. From this plot and the related inclination of $\rm i_{\rm orb}=(102.55 \pm 2.29)^{\circ}$, it is suggested that the trajectory of the binary system is close to edge-on. The size of the blue data points are related to the astrometric uncertainty of $\pm$0.006 as.}
\label{fig:d9_orbit}
\end{figure}

Using the well-known orbit of S2 (S0-2) \cite{Schoedel2002, Do2019S2}, we determine the position of Sgr~A*. Since the intrinsic proper motion of Sgr~A*, v$\rm _{prop, SgrA^*}$, is only a fraction of a pixel per epoch \cite{Parsa2017} and thus several orders of magnitude smaller than the distance to D9, we neglect this velocity term. The rejection of v$_{\rm prop, SgrA^*}$ is motivated by the typical astrometric uncertainties of $\pm 12.5\,$mas that exceed the intrinsic proper motion of Sgr~A* with v$\rm _{prop, SgrA^*}=0.3\,\mathrm{mas/yr}$. From the fixed position of Sgr~A*, we use the astrometric information of D9 to derive a related Keplerian orbital solution. We incorporate the LOS velocity of D9 using the estimated baseline of about $150\,$km/s and a corresponding uncertainty range of $\pm 15\,$km/s. Comparing the statistical significance of the Keplerian fit with and without the LOS velocity results in a difference of almost one magnitude for the reduced $\chi^2$. We estimate $\chi^2_{\rm red}$ to be about 10 for the sole astrometric measurements while we find a robust fit for $\chi^2_{\rm red}$ of approximately 2 by maximizing the parameter space, that is, including the LOS velocity. 
With a mass of M$_{\rm SgrA^*}\,=\,4\times 10^6\,\mathrm{M_{\odot}}$ for Sgr~A* \citep{eht2022, Peissker2022}, we display the resulting Keplerian orbit in Fig. \ref{fig:d9_orbit} and list the corresponding orbital elements in Table \ref{tab:d9_parameter}.
As is evident from the plot displayed in Fig. \ref{fig:d9_orbit}, D9 moves on the descending part of its Keplerian orbit, which results in the mentioned slow velocity. Intriguingly, the relative location and its intrinsic velocity of D9 with respect to Sgr~A* ensure a confusion-free detection of the binary system. Most likely, detection of the binary would be hindered if it was in its ascending part of the orbit.

\subsection*{Statistical analysis}

The Limited-memory Broyden, Fletcher, Goldfarb, and Shannon box constraints (L-BFGS-B) algorithm forms the basis of the Keplerian fit \citep{Liu1989, Zhu1997}. The Keplerian fit relies on the L-BFGS-B algorithm, which is an iterative method that identifies free parameters within a given range and aims to minimize the gap between the data points and the priors (i.e., initial guess). The Keplerian equations of motion describe the model underlying the algorithm. The algorithm iteratively finds the orbital solution that best fits the data points with high accuracy, i.e., the minimized $\chi^2$.\newline
The best-fit parameters are then used as a prior for the Markow-Chain-Monte-Carlo (MCMC) simulations. {The MCMC algorithm was used by the implementation of the emcee PYTHON package developed by \citep{Foreman-Mackey2013}.}
When inspecting the distribution of the measured data points, it is evident that the D9 system moves with a comparable slow velocity in the S cluster, which translates into an almost (projected) linear motion. Hence, it is not entirely unexpected that the MCMC simulations are in high agreement with the best-fit results of the Keplerian approximation (Table \ref{tab:d9_kepler_comp}).
We can conclude that the orbital solution presented in Table \ref{tab:d9_kepler_comp} is robust and should provide a suitable basis for future high-angular resolution observations.

\subsection*{Uniqueness of the IFU data points}

\begin{figure}[ht!]%
\centering
\includegraphics[width=1.0\textwidth]{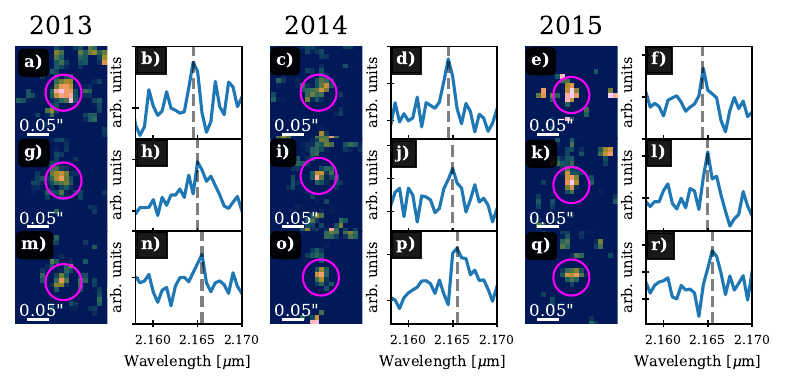}
\caption{{\bf Doppler-shifted Br$\gamma$ line of D9 and the related line maps representing the magenta-marked emission.} Subplots (a), (c), (e), (g), (i), (k), (m), (o), and (q) show SINFONI line maps of the binary system D9. In these subplots, D9 is marked with a magenta-colored circle. In sublots (b), (d), (f), (h), (j), (l), (n), (p), and (q), we apply a local background subtraction of the surrounding gas to the presented spectra. The successful subtraction of the background is evident in the absence of the prominent Br$\gamma$ peak at $2.1661\mu m$ \citep{Peissker2020c, Ciurlo2021}. The shown spectra shows the evolution of the line over one year. The normalized Br$\gamma$ velocity v$_{\rm norm}$ in 2013 is approximately 66 km/s (b), 3 km/s (h), and -72 km/s (n). In 2014, v$_{\rm norm}$ is about 68 km/s (d), 3 km/s (j), and -71 km/s (p). In 2015, we estimate v$_{\rm norm}$ to be around 72 km/s (f), 1 km/s (l), and -67 km/s (r).}
\label{fig:d9_overview}
\end{figure}

The line maps of the three-dimensional data cubes observed with SINFONI and ERIS act as a response actor, which is interpreted as a measure of the influence of nearby sources and the imprint of the background. It is important to note that sporadic background fluctuations do not result in a line map emission counterpart. In other words, the line emission with spatially limited origin (i.e. noise) does not produce a (compact) line map signal comparable to, e.g., G2 \citep{peissker2021c}. This is due to the flux required to produce a signal above the sensitivity level of the detector. Vice versa, only spatially extended emission with sufficient line emission produces a spectroscopic signal (Supplementary Fig. 4 and Supplementary Fig. 5). This interplay between line emission and line maps reduces the chance of detecting false positives of any kind. Mathematically speaking, the mentioned interplay between the two parameter spaces (spatial and spectroscopic) of detecting a real signal is a necessary condition. In this sense, one cannot claim the existence of a source based on one parameter space.\newline
Taking into account the Keplerian orbit of D9 further reduces the probability of a false positive, which occurs only at the expected orbit position, by several magnitudes. \cite{Sabha2012} and \cite{Eckart2013} calculated the probability of detecting an artificial source on a Keplerian orbit to be in the range of a fraction of a percent. This can only be considered an upper limit because the probability relates to a time span of 5 years and covers solely astrometric data. 
In Fig. \ref{fig:d9_overview}, we show an overview of selected epochs to demonstrate the interplay between the observed Br$\gamma$ emission line and the line maps. These line maps are created by selecting a wavelength range of about $0.0015\mu m$, which corresponds to three channels in total (out of 2172 channels in total).
A crucial pillar of the binary detection presented in this work is the analysis of individual nights observed with SINFONI and ERIS. Therefore, it is expected that the quality of the data will differ not only due to variable weather conditions but also to the number (i.e., on-source integration time) of observations executed at the telescope (Fig. \ref{fig:d9_overview}). Of course, the impact of these boundary conditions is reduced by stacking individual cubes, as has been done for the analysis presented, for example, in \citep{Peissker2019, Peissker2020b, peissker2021c}. Since the RV signal of the D9 system changes on a daily basis, stacking these single night data cubes affects the signal-to-noise ratio (SNR) of the Br$\gamma$ line emission of the D9 system (Supplementary Fig. 4). For example, the signal-to-noise ratio for the stacked 2019 SINFONI data cube with an on-source integration time of almost 10 hours is 20, while two cubes from a single night in 2019.43 show an average SNR of {about} 5. Although detection of the D9 binary system would benefit from using the data cubes that include all annual observations, an analysis of the periodic RV signal would be hindered.

\subsection*{ERIS data}

\begin{figure}[ht!]%
\centering
\includegraphics[width=0.8\textwidth]{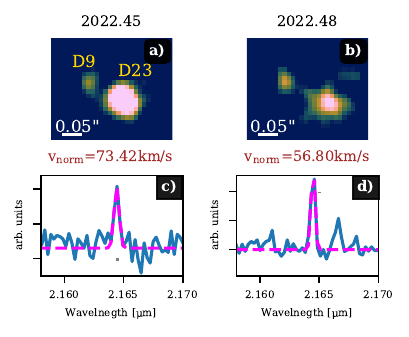}
\caption{{\bf Observations of the D9 binary system in 2022 with ERIS.} In subplots (a) and (b), the Br$\gamma$ line maps observed with ERIS in 2022 are shown. Both subplots display the binary system D9 and the close-by source D23. For visualization purposes, we apply a 40 mas Gaussian kernel to these line maps. Subplots (c) and (d) show the related spectrum where we indicate the normalized RV velocity v$_{\rm norm}$. Including the offset measured by Exo-Striker of about 29 km/s, these velocities are displayed as black data points in Fig. \ref{fig:d9_rv}.}
\label{fig:eris_data}
\end{figure}

The ERIS data analyzed in this work are part of the science verification observations carried out in 2022 by the PI team. To reduce the data, we use the ESO pipeline \citep{Freudling2013} that applies the standard procedure (dark, flat, and distortion correction). Furthermore, the data are part of a preliminary analysis of the Galactic center with ERIS \citep{Davies2023}. The authors of \cite{Davies2023} report a superior performance compared to SINFONI, which can be confirmed as shown in Fig. \ref{fig:eris_data}.
Although the on-source integration time is only 1200 seconds for each night, we find an SNR of almost 6 for the Doppler-shifted Br$\gamma$ emission line of the D9 binary system. In both data sets shown, we detect D9 close to D23 without confusion comparable to the SINFONI observations displayed in Fig. \ref{fig:d9_multiwavelength_detection} at the expected wavelength (Fig. \ref{fig:d9_rv}). Due to the distance between D9 and D23 in 2022, both sources will be affected by interference in forthcoming observations of the S~cluster.

\subsection*{Radial velocity fit}
 
For the spectrum that is used to extract the related LOS velocity, we subtract the underlying continuum by fitting a polynomial to the spectroscopic data. Line maps are constructed in the same way directly from the three-dimensional data cubes (Fig. \ref{fig:d9_multiwavelength_detection}). Using an aperture with a radius of 25 mas, the extracted spectrum of D9 reveals a velocity range between $\rm -67\,$km/s and $\rm -225\,$km/s (Supplementary Tables 4-6) on the investigated data baseline with a corresponding average LOS velocity of v$\rm _{\rm LOS}\,=\,-153.72\,$km/s and a measured uncertainty of $\rm 16.38\,$km/s (Table \ref{tab:d9_parameter}). If the source is isolated, we use an annulus for a local background subtraction \citep{Valencia-S.2015}. In any other case, we select an empty region 0.1" west of S59 (Fig. \ref{fig:d9_multiwavelength_detection}). Subtracting the baseline (v$_{\rm min}$+v$_{\rm max}$)/2 from the individual velocity values normalizes the distribution. With this arrangement of the observed RV, we used the tool Exo-Striker \cite{exostriker2019} to fit the related velocities, which resulted in the binary orbital parameter listed in Table \ref{tab:d9_parameter} and the Keplerian fit of the secondary trajectory displayed in Fig. \ref{fig:d9_rv}. The model predicts a secondary on an elliptical orbit around the primary, which further results in an RV offset of about 29 km/s. This offset is added to the normalized velocities. As shown in Fig. \ref{fig:d9_rv}, the final normalized LOS velocity is around $-120$km/s.
The reduced chi-square is $\rm \chi^2=0.31$, which implies a significant agreement between the data and the fit. Due to the extended data baseline of 15 years (Supplementary Tables 7-9), we established an independent sanity check to reflect the satisfactory agreement of the observed RV and the fit. For this, we split the data and limit the fit to the epochs between 2013 and 2019. Hence, the epochs before 2013 represent a noncorrelated parameter to the Keplerian model provided by Exo-Striker with an average LOS velocity of v$\rm _{\rm LOS*}\,=\,-147\,$km/s. The difference between the average v$\rm _{\rm LOS}$ and v$\rm _{\rm LOS*}$ is expected due to the phase coverage and the intrinsic LOS velocity of D9. We note that both averaged velocities are within the estimated uncertainties. It is also notable that the independent RV data before 2013 and after 2019 match the derived periodic model of the D9 binary system.

\bmhead{Data availability}

The datasets generated during and/or analyzed during the current study are available from the corresponding author upon request.

\bmhead{Code availability}

The code for generating the SED is publicly available at \url{http://www.hyperion-rt.org/}. Stable version 1.4 was used to generate the SED. The evolutionary tracks {PARSEC} can be found at \url{http://stev.oapd.inaf.it/cgi-bin/cmd} (version 3.7). The radial fit was performed with Exo-Stricker, version 0.88, and can be found at \url{https://exo-restart.com/tools/the-exo-striker-tool/}. The emcee package is a pure PYTHON package and can be downloaded from \url{https://emcee.readthedocs.io/en/stable/}. The ESO pipeline can be downloaded from \url{https://www.eso.org/sci/software/pipelines/}.


\bmhead{Acknowledgments} F.P., L.L., and E.B. gratefully acknowledges the Collaborative Research Center 1601 funded by the Deutsche Forschungsgemeinschaft (DFG, German Research Foundation) -- SFB 1601 [sub-project A3] -- 500700252. MZ acknowledges the financial support of the Czech Science Foundation Junior Star grant no. GM24-10599M. VK acknowledges the Czech Science Foundation (ref.\ 21-06825X).

\bmhead{Author Contributions Statement} 

F.P. discovered the binary system, performed most of the analysis, and led the writing of the manuscript. M.Z. provided the HR diagram and was responsible for the analysis and calculation about dynamical processes. L.L., A.E., and V.K. contributed to the interpretation of the data. E.B. provided contributions to the background of binaries close to massive stars. M.M. contributed to the SED analysis. M.Z., E.B., M.M., and V.K. improved the text. All authors contributed to the writing of the manuscript.

\bmhead{Competing Interests Statement} 

The authors declare no competing interests.
\clearpage
\subsection*{Tables}

\begin{table}[hbt!]
    \centering
    \begin{tabular}{|cc|}
         \hline
           \multicolumn{2}{|c|}{Secondary Keplerian Parameter}  \\
         \hline
        $P_{\rm D9b}$ [year]                           &   1.02 $\pm$ 0.01 \\
        e$_{\rm D9b}$                      &   0.45 $\pm$ 0.01 \\
        $\omega_{\rm D9b}$ [deg]           & 311.75 $\pm$ 1.65 \\
        a$_{\rm D9b}$ [au]                 &   1.59 $\pm$ 0.01 \\
        i$_{\rm D9b}$ [deg]                &   90.00           \\
        $m \sin (\mathrm{i}_{\rm D9b})$ [$\mathrm{M_{\odot}}$] &   0.73 \\
        RV$_{\rm off}$   [km\,s$^{-1}$]    & -29.19 $\pm$ 3.00 \\
        $\chi_{\nu}^2$                     &   0.31            \\
        $rms$ [km\,s$^{-1}$]               &  16.38            \\
         \hline
           \multicolumn{2}{|c|}{Keplerian Parameter for D9 orbiting Sgr~A*} \\
         \hline
        $e_{\rm D9a}$            &   0.32   $\pm$ 0.01   \\
        i$_{\rm D9a}$ [deg]      &   102.55 $\pm$ 2.29   \\
        $a_{\rm D9a}$ [mpc]      &   44.00  $\pm$ 2.42   \\
        $\omega_{\rm D9a}$ [deg] &   127.19 $\pm$ 7.50   \\
        $\Omega_{\rm D9a}$ [deg] &   257.25 $\pm$ 1.61   \\
        P$_{\rm D9a}$ [yr]       &   432.62 $\pm$ 0.01 \\
        \hline
           \multicolumn{2}{|c|}{Radiative Transfer Model} \\
        \hline
        i$_{\rm intrinsic}$ [deg]                    & 75.0 $\pm$ 19.0 \\
        R [R$_{\odot}$]              & 2.00 $\pm$ 0.13 \\
        log(L/L$_{\odot}$)           & 1.86 $\pm$ 0.14 \\
        log(T$_{\rm D9a}$[K])                    & 4.07 $\pm$ 0.05 \\
        M$_{\rm D9a}$ [M$_{\odot}$]   & 2.80 $\pm$ 0.50 \\
        M$_{\rm Disk}$ [10$^{-6}$ M$_{\odot}$ ] & 1.61 $\pm$ 0.02  \\
        \hline
    \end{tabular}
    \caption{{\bf Best-fit parameters of the D9 system.} We list the orbital parameters for the binary of D9 together with the motion of the system around Sgr~A*. In addition, the best-fit stellar properties based on the SED fitter are included. The uncertainties of the binary parameter and the radiation transfer model are based on the reduced $\chi^2$. For the Keplerian elements, we use MCMC simulations to estimate the uncertainty range. {Since the inclination of the secondary is assumed to be $\mathrm{i}_{\rm D9b}\,=\,90^{\circ}$, no uncertainty for $m \sin (\mathrm{i}_{\rm D9b})\,=\,0.73$~M$_{\odot}$ is given.}}
    \label{tab:d9_parameter}
\end{table}

\begin{table}[hbt!]
    \centering
    \begin{tabular}{|cccc|}
         \hline
           Parameter & Best-fit & MCMC & {Standard deviation} \\
         \hline
        a$_{\rm D9a}$ [mpc]                   &   44.00   & 45.55    &  1.15  \\
        e$_{\rm D9a}$                         &   0.32    & 0.31     &  0.01   \\
        i$_{\rm D9a}$ [$^\circ$]            &   102.55  & 103.30   &  1.14   \\
        $\omega_{\rm D9a}$ [$^\circ$]       &   127.19  & 130.96   &  8.02   \\
        $\Omega_{\rm D9a}$ [$^\circ$]       &   257.25  & 258.40   &  1.71   \\
        $t_{\rm closest}$ [years] &   2309.13 & 2315.83  &  7.01   \\
         \hline
    \end{tabular}
    \caption{{\bf Comparison of best-fit Keplerian approximation and MCMC simulations.} Since the standard deviation does not satisfactorily reflect the astrometric precision that can be achieved with SINFONI, we will use the standard deviation of the combined MCMC posteriors. {These orbital elements are related to the outer binary system D9-Sgr~A*.} We refer to \cite{peissker2021c} for a detailed explanation of the background fluctuations of the SINFONI data.}
    \label{tab:d9_kepler_comp}
\end{table}

\clearpage

\section*{Supplementary Information}

\clearpage

\subsection*{Image sharpener}

Filtering techniques are a common tool to enhance the amount of information that can be extracted from astrophysical data. For example, the Keplerian approximation presented in \cite{Schoedel2002} for the B2V star S2/S-02 is based on high-pass filtered data observed with the NTT telescope. In addition, the comprehensive list of S-stars provided by \cite{Gillessen2009} used a high-pass filtering technique. For the photometric analysis presented in this work, we applied a robust image sharpener to the data to filter the high spatial frequencies associated with noise and destructive PSF features that hinder the detection of faint sources. For this, we apply a PSF-sized smoothing kernel to the data I$_{\rm data}$ to obtain a blurred version I$_{\rm smooth}$ of the input file. By subtracting I$_{\rm smooth}$ from I$_{\rm data}$, we get the sharpened version of I$_{\rm data}$. For visual improvements, the final sharpened output can be convolved again with a Gaussian kernel, as was done for the multi-wavelength detection of D9 presented in this work. The astrometric and photometric robustness of this method is demonstrated in \cite{peissker2023c}, where the impact of the image sharpener on all stars observed in the NSC has been analyzed in detail using NACO data. For this work, we use the H+K SINFONI data and compare the filtered and non-filtered data. We apply a 6 pixel (75 mas) Gaussian to the data displayed in the finding chart of D9 and detect in total 53 sources in the non-filtered H and K band data. In the filtered data, we detect about 20$\%$ more sources as a result of the reduced impact of the PSF wings of the stars, which is expected if we compare the total number of S stars reported in \cite{Eisenhauer2005} and \cite{Ali2020}. Eisenhauer et al. used continuum data and reported about 30 S stars. In comparison, Ali et al. used high-pass filtered data and showed approximately $20\%$ more stars inside the inner 40 mpc.
\begin{figure}[ht!]%
\centering
\includegraphics[width=.8\textwidth]{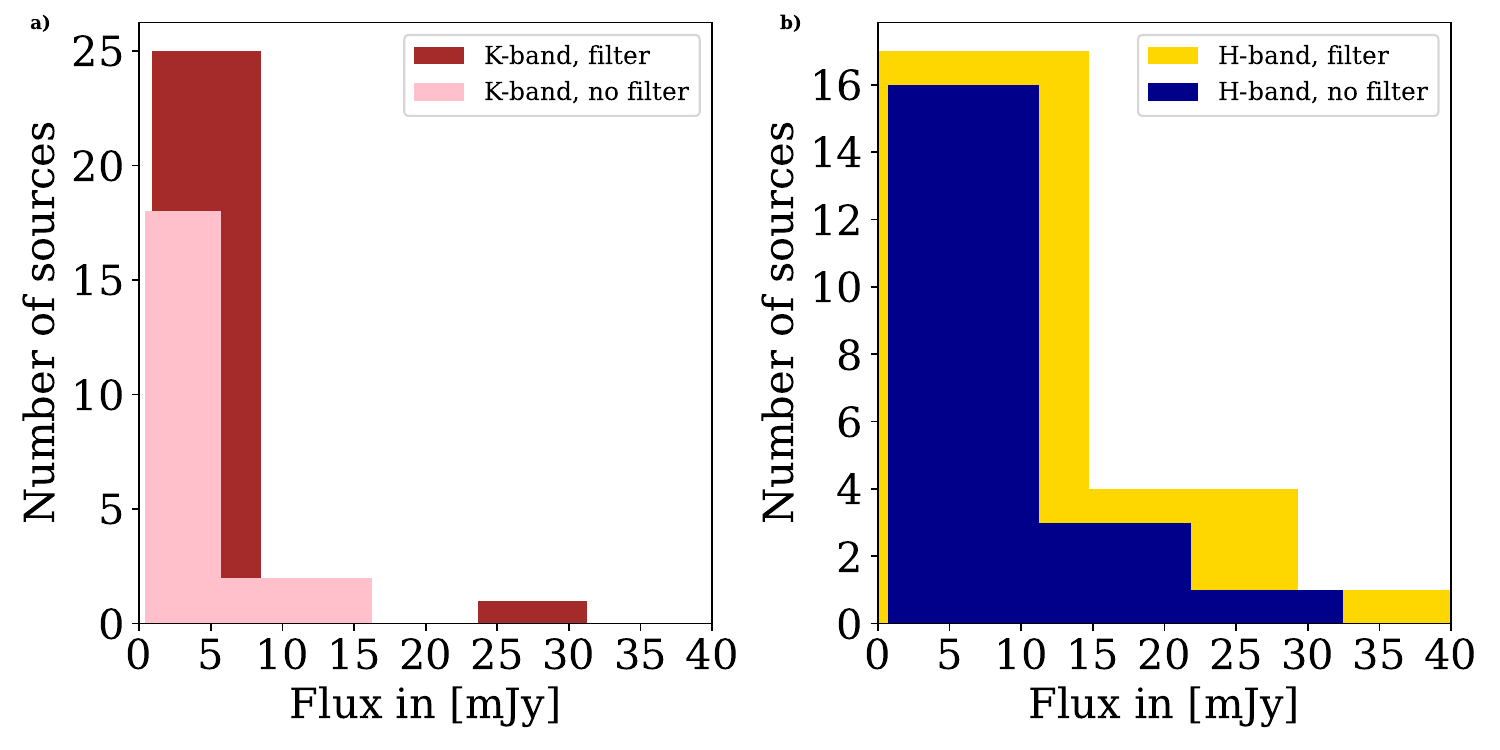}
\caption{{\bf Robustness of the filtering technique used in this work.} We compare the flux of single S-stars in the H- and K-band for filtered and non-filtered data. In subplot (a), we show the K band, whereas subplot (b) represents the H band data. Due to the nature of the image sharpener, we detect about 20$\%$ more individual sources in the filtered data because of the reduced impact of the PSF wings. The bin size in the histograms is half of the square root of the number of sources.}
\label{fig:filter_comparison}
\end{figure}
As a qualitative measure, we compare the median filtered and non-filtered flux density values for the related infrared bands (Fig. \ref{fig:filter_comparison}). We list the resulting flux density values in Table \ref{tab:flux_filter_comp} and find deviations between the filtered and non-filtered data in the range of about $2-3\%$. 
\begin{table}[hbt!]
    \centering
    \begin{tabular}{|ccc|}
         \hline
                                      & H band [mJy]   & K band [mJy]   \\
         \hline
        Non-filtered                  &  5.49  &  3.36  \\
        Filtered                      &  5.78  &  3.54  \\
        Standard deviation            &  0.14  &  0.09  \\
         \hline
    \end{tabular}
    \caption{{\bf Median flux density values for the S stars.} The standard deviation is estimated from the flux values.}
    \label{tab:flux_filter_comp}
\end{table}
Taking into account the expected deviations that are associated with observations of the Galactic center of approximately $10-20\%$ \citep{Ott1999,Viehmann2006,peissker2023c}, the application of an image sharpener does not alter the photometric results analyzed and discussed in this work. Furthermore, the derived flux uncertainties are similar to the estimates used and discussed in \cite{Gautam2024}.

\subsection*{Photometric analysis}

For the photometric analysis of D9, we use the B2V star S2 as the reference source with related dereddened magnitudes of m$\rm _H=$15.9$\pm$0.1, m$\rm _K=$14.1$\pm$0.1, m$\rm _L=$12.6$\pm$0.7 \citep{Peissker2024}. In the continuum data, the D9 system faces challenging confusion effects from nearby stars, such as S2, S12, and S39, so we focus on later epochs to detect the isolated H, K, and L band emission associated with the Doppler-shifted Br$\gamma$ line emission. To minimize the impact of overlapping PSF wings in the crowded S~cluster, we subtract a Gaussian-smoothed version I$_{\rm smooth}$ of the input data I$_{\rm in}$. This post-processing technique is suitable for a wide range of astrophysical objects \cite{Labadie2011} and is characterized by its robustness with respect to image information conservation \cite{peissker2023c}. We use I$_{\rm in}$-I$_{\rm smooth}$ to obtain I$_{\rm out}$ which is shown in the finding chart of D9 that represents a combination of H+K SINFONI data and NIRC2 (L band) observations carried out in 2019. 
\begin{figure}[h!]%
\centering
\includegraphics[width=1.\textwidth]{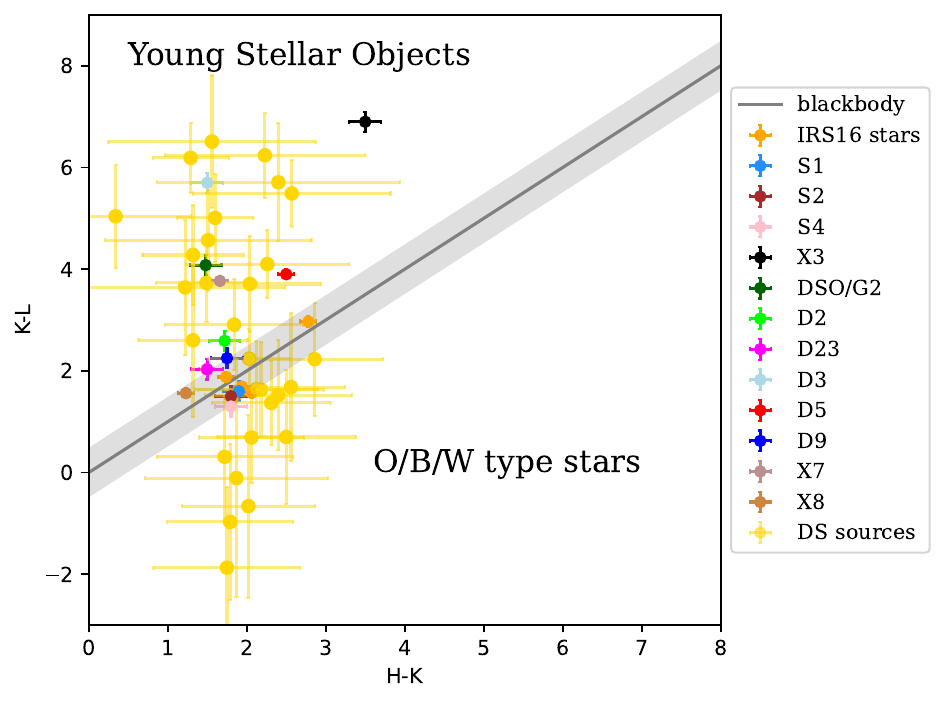}
\caption{{\bf Color-color diagram of prominent sources of the NSC.} In this figure, the linear gray fit represents a blackbody. The young B-stars S1, S2, and S4 are below this linear blackbody fit. All G-objects are above this fit implying a different nature as the main sequence stars of the S~cluster. For comparison, the yellow data points are taken from \cite{peissker2023c} and show the dusty sources in the IRS 13 cluster.}
\label{fig:d9_colors}
\end{figure}
For every data set, we identify a continuum source at the position of the NIR Doppler-shifted Br$\gamma$ emission line and recover the corresponding flux information. We estimate 0.8$\pm$0.1 mJy, 0.3$\pm$0.1 mJy, and 0.4$\pm$0.1 mJy in the H, K, and L band, respectively. The respective dereddened NIR and MIR magnitude values are m$\rm _H=19.92$, m$\rm _K=18.17$, and m$\rm _L=15.92$. From the H-K and K-L colors, we compare the D9 system with other known objects in the S~cluster and the NSC (Figure \ref{fig:d9_colors}). While S stars such as S1, S2, and S4 can be classified as B0-B3 stars \cite{Habibi2017} which is reflected in their related H-K and K-L colors, we find photometric similarities between D9 and two close-by sources D2 (alternatively G3) and D23. All the three sources are located very close to the one-component linear fit (solid black line in Fig. \ref{fig:d9_colors}) emphasizing their multi-wavelength nature. The location of D9 in the color-color diagram suggests a source that already shows characteristics of a main sequence star with a weak dust component.

\subsection*{Extinction}

Although the extinction of the reference source S2 was covered in detail in \cite{Schoedel2010} and \cite{Sabha2012}, we explore the impact of different values on the stellar parameters of the D9 system. The extinction law derived from \cite{Fritz2011} and the related magnitudes resulted in the flux density values for the SED presented in this work. As the magnitudes and, consequently, the flux density values estimated for D9 are extinction corrected, a range of dust densities is introduced into the model, which in turn affects the degree of reddening. These various dust densities can be described with visual extinction A$_V$, which is set to 0 in the SED fit presented in this article (because the magnitudes of the reference star are already extinction corrected). To investigate the impact of variable reddening, we use A$_{V1}$=1 and A$_{V2}$=2, which are 2-4 times larger than the usual uncertainties for optical extinction \citep{Fritz2011}. We list all the results in Table \ref{tab:extinction_comp}, where we implemented different optical extinction values. 
\begin{table}[hbt!]
    \centering
    \begin{tabular}{|ccccc|}
         \hline
          A$_V$                            & Radius in [R$_{\odot}$]  & Temperature in log(T[K]) &  M$_{\rm Disk}$ in [M$_{\odot}$] &  Envelope size in [AU] \\
         \hline
          31                               &  2.00 $\pm$ 0.13  &  4.07 $\pm$ 0.05 & 1.61$\times$10$^{-6}$ & 50.43 \\
          32                               &  2.12 $\pm$ 0.33  &  4.10 $\pm$ 0.09 & 7.37$\times$10$^{-5}$                       &  85.84 \\
          33                               &  2.22 $\pm$ 0.42  &  4.12 $\pm$ 0.08 & 2.61$\times$10$^{-3}$                       & 136.20 \\
         \hline
    \end{tabular}
    \caption{{\bf Comparison of the impact of different extinction values.} For all three inspected extinction values, the stellar radius and temperature of D9 are comparable inside the uncertainties calculated by the radiative transfer model.}
    \label{tab:extinction_comp}
\end{table}
As is evident from Table \ref{tab:extinction_comp}, the different extinction values do not significantly impact the stellar radius or the temperature. Since both values are directly related to the position of D9 on the isochrones shown in Fig. 4 in the main text, we find a stellar age estimate of $2.7^{+1.9}_{-0.3}\,\times\,10^6$~yr. In deriving the disk mass and envelope size of the system using HYPERION, we find values that differ by a factor of one to three (Table \ref{tab:extinction_comp}). Although the small envelope (compared to, e.g., X3 \cite{peissker2023b}) is in line with the mass of the shallow circumbinary disk for all the three extinction values, we want to explore a possible impact on the expected thermal emission in the L band. For this, we will use the empirical model of \cite{Whitney2004} to calculate the dust sublimation radius r$_{\rm sub}$ and the outer radius of the envelope r$_{\rm out}$ that still contributes to the detected emission. Assuming a temperature of approximately 1000 K where dust photoevaporates due to stellar winds, we find a sublimation radius of r$_{\rm sub}$ = r$_{\rm star}(\rm T_{\rm dust}/T_{\rm star})^{-2.085}$ = 1.2 AU. Due to the presence of the secondary, r$_{\rm sub}$ = 1.2 AU can be interpreted as a lower limit because we find a matching effective orbital radius of r$_{\rm eff}$ of approximately 1.2 AU. Presumably, the secondary has cleared the inner dust region of the primary as it has been observed for PDS 70 \citep{Keppler2018} or HD 104237 \citep{Garcia2013}. For an upper limit of the envelope size, we use the usual detection limit for dust sources in the GC of around 200 K \citep{Dinh2024}. We estimate an approximate minimum envelope radius of 50 AU and conclude that the L band emission for all the three extinction values should be observable on a comparable level. Due to dust shielding, the largest envelope listed in Table \ref{tab:extinction_comp} will have low temperatures below the NIRC2 detection limit in the L band of 0.04 mJy. The detection limit for colder regions further away from the central star estimated in this work is in agreement with the analysis of \cite{Shimizu2023}. An additional effect of different extinction values on the stellar parameters is the disk mass M$_{\rm Disk}$. However, we should note that the impact on the radius and temperature of D9 is limited, as shown in Table \ref{tab:extinction_comp}. Consequently, we do not expect a significant variation in the mass M$_{\rm D9a}$ of the primary. This is not surprising since a correlation between M$_{\rm Disk}$ and stellar mass has not been observed so far \citep{Stapper2022}. Overall, we find that the extinction used in this work is robust against possible variations on small scales \citep{Schoedel2010, Peissker2020c}.

\subsection*{Instrumental effects}

As mentioned, the data are corrected for the variation of the barycenter of the Earth. However, we exclude the possibility of a false detection due to instrumental effects by investigating the close-by source D23. We use the same data set that resulted in the detection of the binary system D9 for the analysis of the Doppler-shifted Br$\gamma$ emission line of D23.
\begin{figure}[ht!]%
\centering
\includegraphics[width=1.\textwidth]{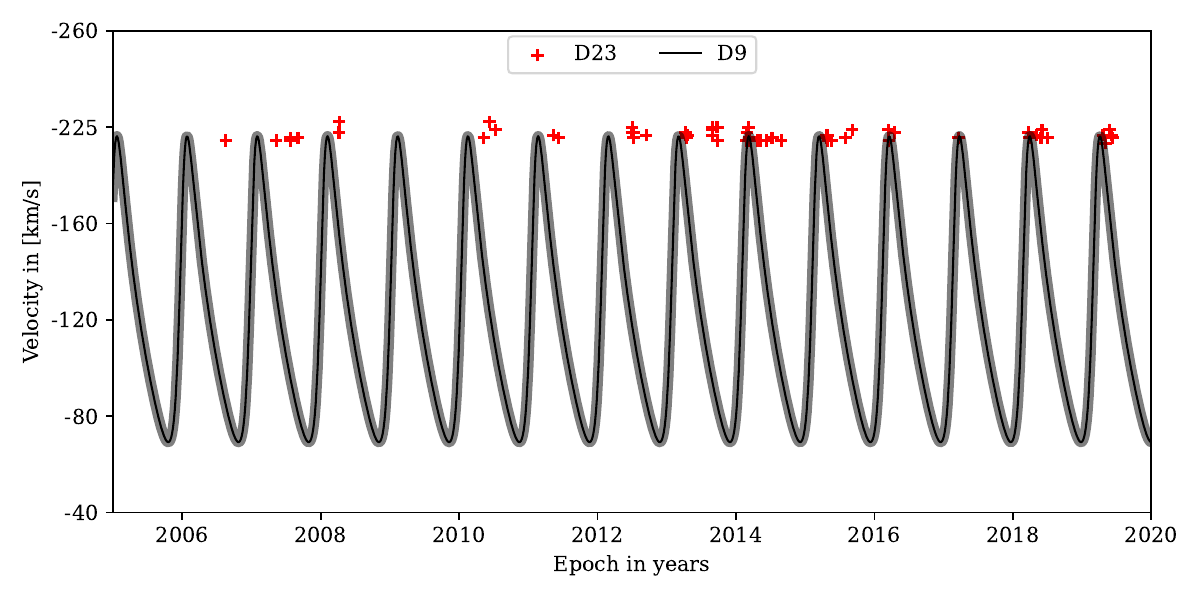}
\caption{{\bf Blue-shifted LOS velocity of D23.} The close-by source with a velocity of $-220.16\pm2.47\,$km/s shows no signs of a periodic signal, which excludes an artificial origin of the RV pattern of D9. The uncertainty indicated reflects the standard deviation of the observed LOS velocities. The gray pattern shows the RV model displayed in the periodic pattern of D9 and demonstrates the magnitude of the LOS variability of D9 as compared to D23.}
\label{fig:d23_d9_comparison}
\end{figure}
Due to the confusion-free detection of D23 between 2005 and 2019, as shown in Fig \ref{fig:d23_d9_comparison}, we conclude that instrumental effects do not cause the periodic RV pattern. Furthermore, the two ERIS data points observed in 2022 show the same pattern as observed with SINFONI. Even if both instruments were to generate a periodic pattern artificially, it is implausible that the magnitudes of these signals would be in agreement. We add that no periodicity is reported for S2 (S0-2) by examining the Br$\gamma$ absorption line with SINFONI \cite{gravity2018}, strengthening our earlier argument.

\subsection*{Data quality}

For the validity of the findings, we explore the quality of the data used to derive the periodic signal of the binary and listed in Tables \ref{tab:data_sinfo1}-\ref{tab:data_sinfo4} by classifying the observations in medium and good. For this classification, we fit a Gaussian to the K band continuum emission of S2 (S0-2) and compare the Full Width Half Maximum (FWHM) of each data set. A high quality is represented by diffraction-limited values close to 6 pixel (75 mas), whereas medium data are close to 7 pixel (87.5 mas). Usually, higher values forbid confusion-free detection of individual S stars in the cluster. In addition to this quality control, we measure the SNR level to explore the influence of quantities beyond the initial quality control.
\begin{figure}[h!]%
\centering
\includegraphics[width=0.8\textwidth]{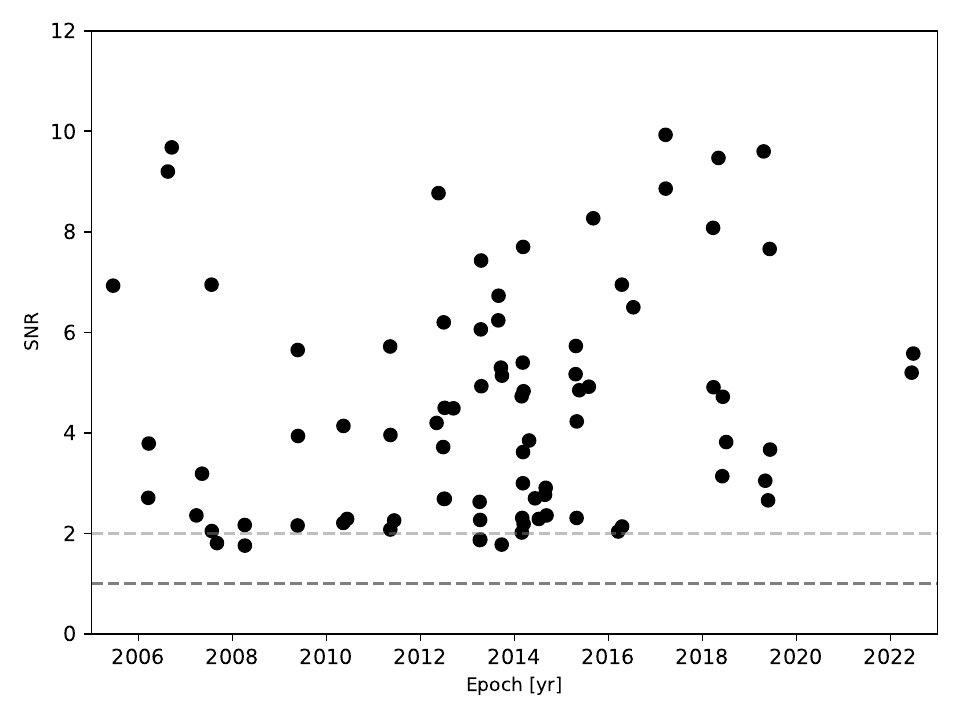}
\caption{{\bf Quality of the data investigated in this work.} Except for five data points, all observations show a SNR above two (indicated with a gray dashed line).}
\label{fig:data_quality}
\end{figure}
For that, we use the maximum of the Br$\gamma$ line and divide it by the noise level of the spectrum as indicated by the dashed fit displayed in the periodic pattern of the binary (Tables \ref{tab:data_sinfo1}-\ref{tab:data_sinfo4}). The final SNR as a function of the epoch is shown in Fig. \ref{fig:data_quality}, where the threshold of 1 represents the noise level (black dashed line). For these data points, find an average SNR of $\rm 4.49\pm2.27$ where the uncertainty represents the standard deviation. While a threshold of SNR=2 is considered to be a non-artificial noise fluctuation, it is important to note that these measurements are executed on the reduced and sky-subtracted data cubes. It is well known that the dark and sky subtraction negatively impacts the SNR of the reduced data \citep{Newberry1991}, which is why we conclude that all the data points used are of a physical nature. This argument is strengthened by the uniqueness of the IFU data points, i.e., all emission peaks show a line map counterpart. In order to visualize the relationship between the line maps and the Doppler-shifted Br$\gamma$ emission peaks, we picked five observations with an SNR below two (Fig. \ref{fig:data_quality}). We display the related line maps together with the corresponding Br$\gamma$ emission with the lowest SNR in the investigated data set in Fig. \ref{fig:D9_low_data_quality}.
\begin{figure*}
\centering
\includegraphics[width=1.0\textwidth]{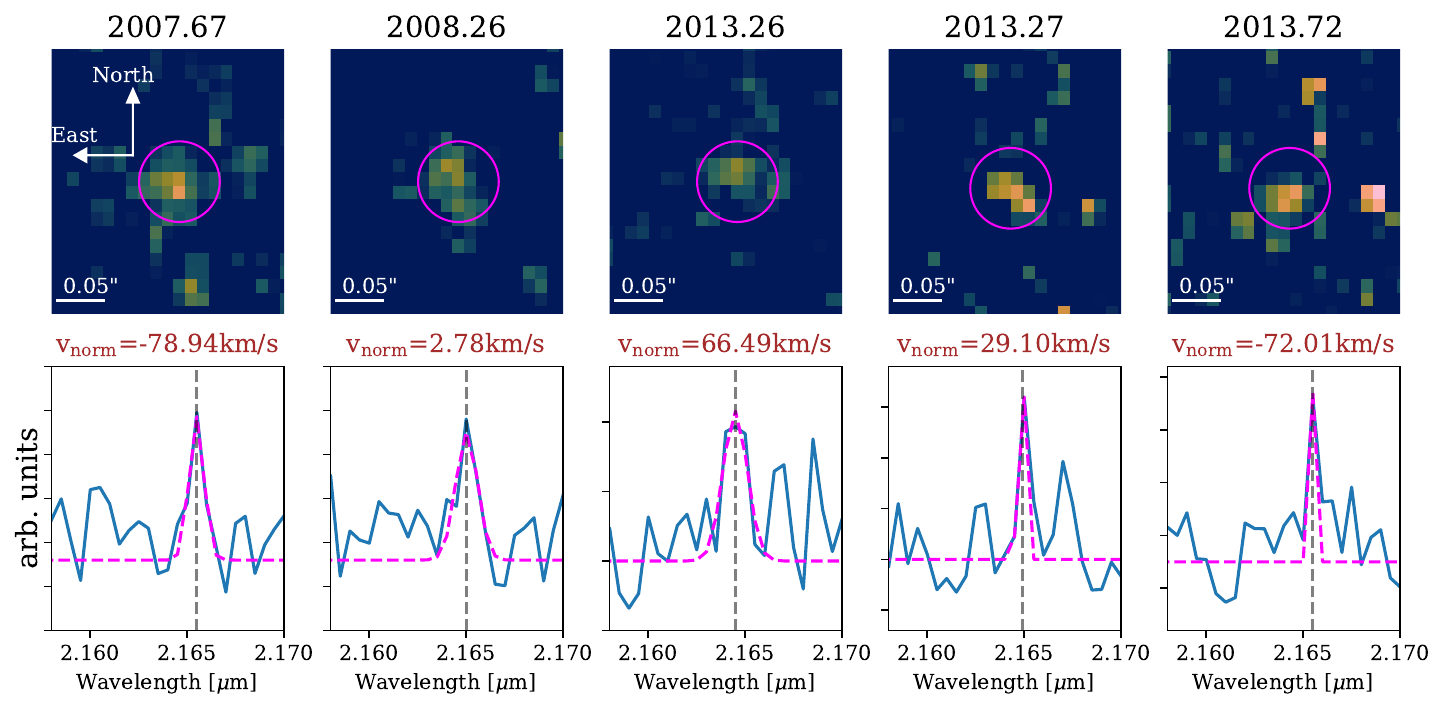}
\caption{{\bf Detection of the binary system D9 in the data set suffering from a low SNR.} The selected data in this figure exhibit the highest noise level compared to any other data displayed in this work. However, the Br$\gamma$ line is Doppler-shifted, which is why the emission of D9 can be distinguished from the sporadic noise in the data. D9 is in the center of each plot, north is up, east is to the left.}
\label{fig:D9_low_data_quality}
\end{figure*}
Despite the low data quality, we detect D9 at the expected astrometric positions with an RV that matches the binary period presented in this work. The line maps in Fig. \ref{fig:D9_low_data_quality} exhibit emission above the noise and are similar to the detection of D9 in the high SNR data.

\newpage

\section*{Supplementary Tables}

\begin{table}[hbt!]
    \centering
    \begin{tabular}{|ccc|}
         \hline
           ID & Period [days] & Reference \\
         \hline
           S2-36         & 39.42  &  \cite{Gautam2019}   \\
           S4-258 (E60)  & 2.30   &  \cite{Pfuhl2014}    \\
           S4-308        & 1.33   &  \cite{Gautam2024}    \\
           IRS 16SW      & 19.45  &  \cite{Ott1999}      \\
           IRS 16NE      & 224.0  &  \cite{Pfuhl2014}     \\
           \hline
           D9      & 372.3  &  This work     \\
        \hline
    \end{tabular}
    \caption{{\bf Binaries of the Galactic center.} In this list, D9 is the only S~cluster member.}
    \label{tab:binary_overview}
\end{table}

\newpage
\section*{Supplementary Note 1}
\newpage

\begin{figure}
\centering
\includegraphics[width=1.1\textwidth]{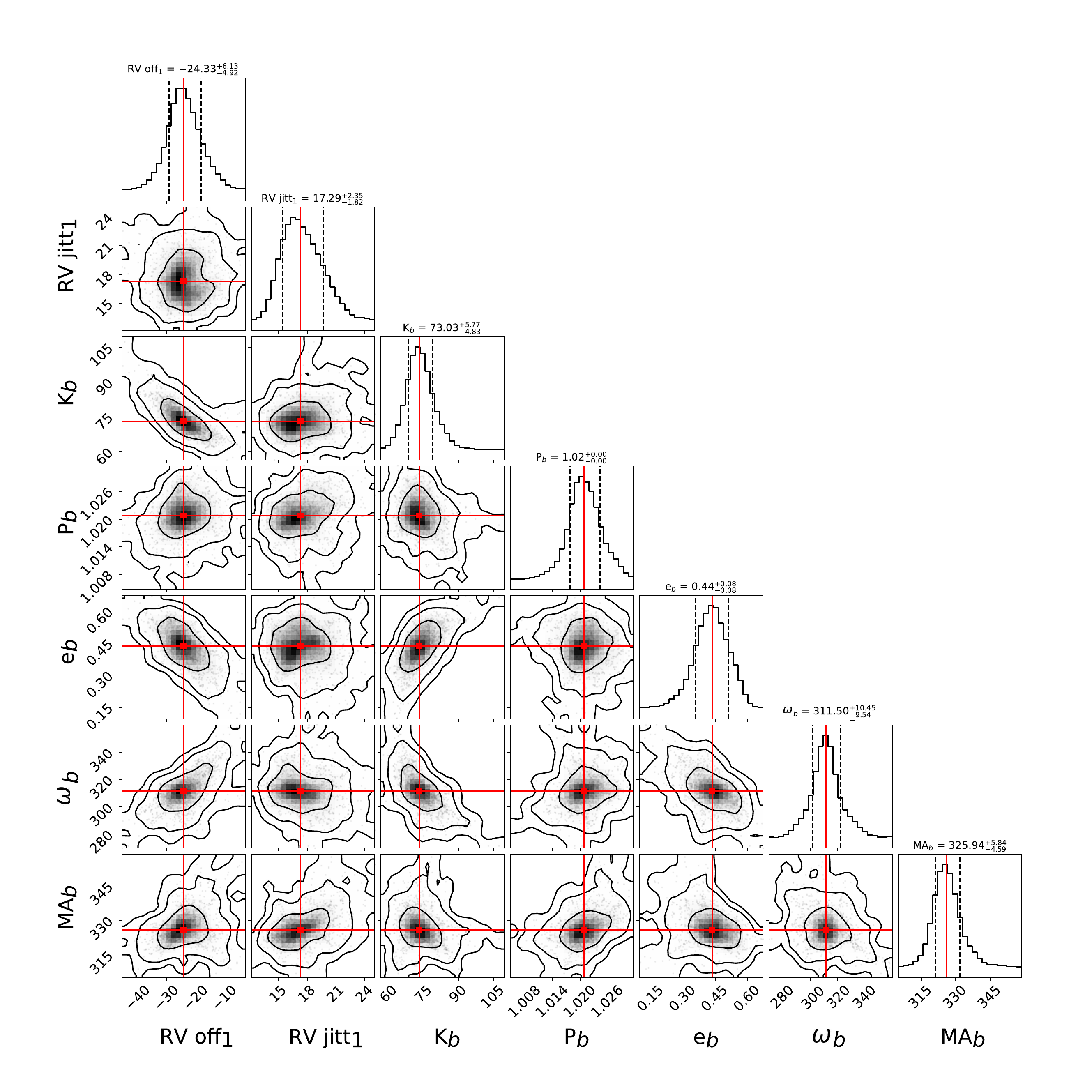}
\caption{{\bf Cornerplot of the Keplerian orbital parameters describing the secondary of the D9 binary system.} We use the fit results of Exo-Stricker as priors to the MCMC model. The model is in high agreement with the fit results.}
\label{fig:cornerplot}
\end{figure}

\newpage

\begin{figure}
\centering
\includegraphics[width=0.8\textwidth]{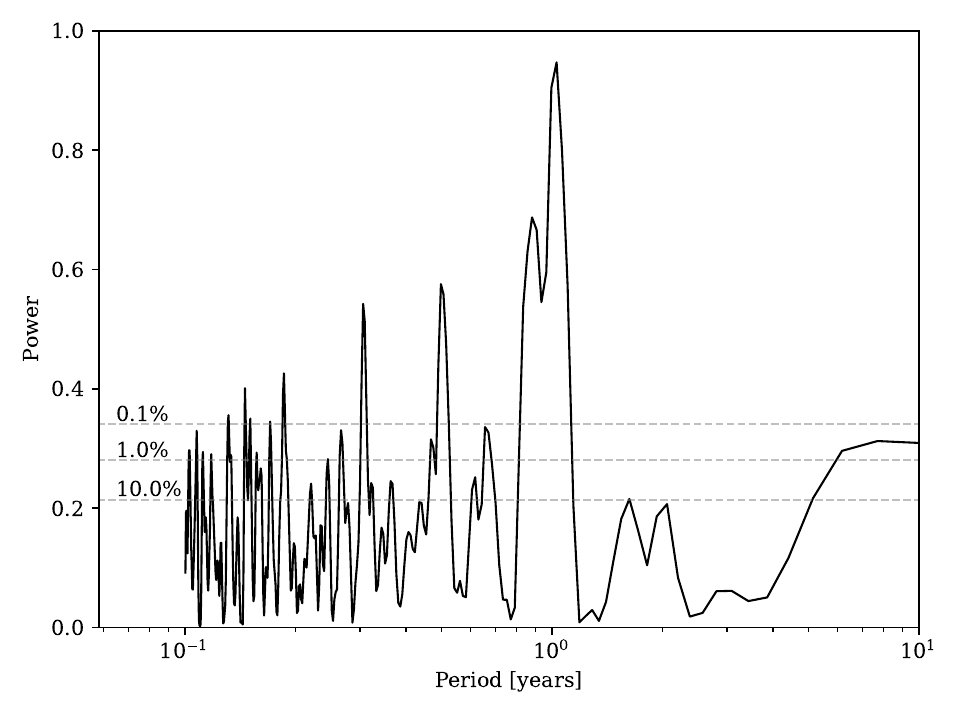}
\caption{{\bf Generalized Lomb-Scargle diagram of the frequency of the period.} The x-axis is logarithmically scaled. The dashed lines indicate the False Alarm Probability (FAP) with 10.0$\%$, 1.0$\%$, and 0.1$\%$. All peaks above are statistically significant. The peak maximum shows the LS power of 0.94 and can be found at 1.02 years.}
\label{fig:lombscargle}
\end{figure}

\newpage

\begin{table*}
    \centering
    \begin{tabular}{|ccccccc|}
         \hline
           Epoch & Wavelength [$\mathrm{\mu m}$] & FWHM [$\mathrm{\mu m}$] & RV [km/s] & $\Delta$RV [km/s] & v$_{\rm norm}$ [km/s] & v$_{\rm mod}$ [km/s] \\
         \hline 
2005.463 & 2.16522 & 0.00035 & -121.87 & 2.76 & -24.92 &    4.27 \\
2006.210 & 2.16498 & 0.00038 & -155.11 & 1.38 &   8.30 &   37.49 \\
2006.221 & 2.16499 & 0.00089 & -153.73 & 8.30 &   6.92 &   36.11 \\
2006.627 & 2.16549 & 0.00080 &  -84.48 & 1.38 & -62.32 &  -33.12 \\
2006.710 & 2.16546 & 0.00075 &  -88.63 & 2.76 & -58.16 &  -28.96 \\
2007.235 & 2.16496 & 0.00142 & -157.88 & 9.69 &  11.07 &   40.26 \\
2007.355 & 2.16501 & 0.00091 & -150.96 & 4.15 &   4.15 &   33.34 \\
2007.557 & 2.16549 & 0.00088 &  -84.48 & 2.76 & -62.32 &  -33.12 \\
2007.562 & 2.16548 & 0.00089 &  -85.86 & 5.53 & -60.93 &  -31.74 \\
2007.672 & 2.16561 & 0.00073 &  -67.86 & 11.07& -78.94 &  -49.75 \\
2008.263 & 2.16489 & 0.00091 & -167.58 & 9.69 &  20.77 &   49.96 \\
2008.266 & 2.16500 & 0.00098 & -152.34 & 1.38 &   5.53 &   34.72 \\
2009.388 & 2.16541 & 0.00104 &  -95.56 & 5.53 & -51.24 &  -22.05 \\
2009.390 & 2.16548 & 0.00103 &  -85.86 & 2.76 & -60.93 &  -31.74 \\
2009.396 & 2.16539 & 0.00062 &  -98.33 & 9.69 & -48.47 &  -19.27 \\
2010.360 & 2.16487 & 0.00128 & -170.35 & 9.69 &  23.54 &   52.73 \\
2010.363 & 2.16489 & 0.00076 & -167.58 & 11.07&  20.77 &   49.96 \\
2010.441 & 2.16500 & 0.00058 & -152.34 & 1.38 &   5.53 &   34.72 \\
2011.357 & 2.16498 & 0.00142 & -155.11 & 2.76 &   8.30 &   37.49 \\
2011.360 & 2.16501 & 0.00096 & -150.96 & 4.15 &   4.15 &   33.34 \\
2011.363 & 2.16501 & 0.00092 & -150.96 & 1.38 &   4.15 &   33.34 \\
2011.441 & 2.16497 & 0.00091 & -156.50 & 4.15 &   9.69 &   38.88 \\
2012.344 & 2.16458 & 0.00076 & -210.51 & 6.92 &  63.70 &   92.89 \\
2012.385 & 2.16498 & 0.00091 & -155.11 & 2.76 &   8.30 &   37.49 \\
2012.485 & 2.16502 & 0.00060 & -149.57 & 2.76 &   2.76 &   31.95 \\
2012.496 & 2.16499 & 0.00089 & -153.73 & 5.53 &   6.92 &   36.11 \\
2012.500 & 2.16501 & 0.00091 & -150.96 & 1.38 &   4.15 &   33.34 \\
2012.516 & 2.16500 & 0.00065 & -152.34 & 1.38 &   5.53 &   34.72 \\
2012.519 & 2.16551 & 0.00087 &  -81.71 & 4.15 & -65.09 &  -35.90 \\
2012.702 & 2.16550 & 0.00093 &  -83.09 & 1.38 & -63.70 &  -34.51 \\
        \hline
    \end{tabular}
    \caption{{\bf Radial velocities measured from the SINFONI observations carried out between 2005-2012.} We list the epoch, the wavelength, the corresponding Full-Width-Half-Maximum, the LOS velocity, and the related uncertainty. Furthermore, we indicate the normalized RV v$_{\rm norm}$ and the final velocity v$_{\rm mod}$ where the Keplerian model for the secondary from Exo-Stricker is included. All velocities, except v$_{\rm mod}$, are estimated with respect to the Br$\gamma$ rest wavelength at $2.1661\,\mathrm{\mu m}$.}
    \label{tab:rv_values_1}
\end{table*}

\newpage

\begin{table*}
    \centering
    \begin{tabular}{|ccccccc|}
         \hline
           Epoch & Wavelength [$\mathrm{\mu m}$] & FWHM [$\mathrm{\mu m}$] & RV [km/s] & $\Delta$RV [km/s] & v$_{\rm norm}$ [km/s] & v$_{\rm mod}$ [km/s] \\
         \hline
2013.260 & 2.16451 & 0.00125 & -220.21 & 5.53 &  66.47 &   95.66 \\
2013.263 & 2.16477 & 0.00113 & -184.20 & 6.92 &  30.46 &   59.65 \\
2013.271 & 2.16478 & 0.00080 & -182.81 & 1.38 &  29.08 &   58.26 \\
2013.274 & 2.16502 & 0.00070 & -149.57 & 2.76 &  -4.15 &   25.04 \\
2013.285 & 2.16492 & 0.00108 & -163.42 & 4.25 &   9.69 &   38.88 \\
2013.291 & 2.16496 & 0.00116 & -157.88 & 6.92 &   4.15 &   33.34 \\
2013.296 & 2.16497 & 0.00076 & -156.50 & 1.38 &   2.76 &   31.95 \\
2013.657 & 2.16550 & 0.00090 &  -83.09 & 4.15 & -70.63 &  -41.44 \\
2013.662 & 2.16550 & 0.00070 &  -83.09 & 1.38 & -70.63 &  -41.44 \\
2013.713 & 2.16548 & 0.00105 &  -85.86 & 2.76 & -67.86 &  -38.67 \\
2013.729 & 2.16551 & 0.00072 &  -81.71 & 1.38 & -72.01 &  -42.82 \\
2013.735 & 2.16548 & 0.00099 &  -85.86 & 1.38 & -67.86 &  -38.67 \\
2014.154 & 2.16448 & 0.00117 & -224.36 & 2.76 &  70.63 &   99.82 \\
2014.157 & 2.16448 & 0.00122 & -224.36 & 2.76 &  70.63 &   99.82 \\
2014.166 & 2.16450 & 0.00100 & -221.59 & 1.38 &  67.86 &   97.05 \\
2014.177 & 2.16452 & 0.00078 & -218.82 & 6.92 &  65.09 &   94.28 \\
2014.180 & 2.16450 & 0.00089 & -221.59 & 2.76 &  67.86 &   97.05 \\
2014.183 & 2.16452 & 0.00084 & -218.82 & 1.38 &  65.09 &   94.28 \\
2014.185 & 2.16450 & 0.00081 & -221.59 & 2.76 &  67.86 &   97.05 \\
2014.194 & 2.16452 & 0.00075 & -218.82 & 2.76 &  65.09 &   94.28 \\
2014.196 & 2.16450 & 0.00073 & -221.59 & 1.38 &  67.86 &   97.05 \\
2014.312 & 2.16477 & 0.00127 & -184.20 & 6.92 &  30.46 &   59.65 \\
2014.438 & 2.16497 & 0.00118 & -156.50 & 1.38 &   2.76 &   31.95 \\
2014.519 & 2.16507 & 0.00104 & -142.65 & 2.76 & -11.07 &   18.12 \\
2014.651 & 2.16543 & 0.00110 &  -92.79 & 8.30 & -60.93 &  -31.74 \\
2014.665 & 2.16550 & 0.00126 & -83.098 & 1.38 & -70.63 &  -41.44 \\
2014.683 & 2.16555 & 0.00107 & -76.173 & 1.38 & -77.54 &  -48.35 \\ 
        \hline
    \end{tabular}
    \caption{{\bf Radial velocities measured from the SINFONI observations carried out between 2013-2014.} Like Table \ref{tab:rv_values_1}, we list v$_{\rm mod}$ which is the final output velocity from the RV fitter Exo-Stricker.}
    \label{tab:rv_values_2}
\end{table*}

\newpage

\begin{table*}
    \centering
    \begin{tabular}{|ccccccc|}
         \hline
           Epoch & Wavelength [$\mathrm{\mu m}$] & FWHM [$\mathrm{\mu m}$] & RV [km/s] & $\Delta$RV [km/s] & v$_{\rm norm}$ [km/s] & v$_{\rm mod}$ [km/s] \\
         \hline
2015.302 & 2.16447 & 0.00108 & -225.75 & 4.15 &  72.01 &  101.20 \\
2015.307 & 2.16451 & 0.00088 & -220.21 & 1.38 &  66.47 &   95.66 \\
2015.323 & 2.16452 & 0.00109 & -218.82 & 2.76 &  65.09 &   94.28 \\
2015.326 & 2.16498 & 0.00132 & -155.11 & 4.15 &   1.38 &   30.57 \\
2015.379 & 2.16504 & 0.00106 & -146.80 & 5.53 &  -6.92 &   22.27 \\
2015.583 & 2.16513 & 0.00119 & -134.34 & 6.92 & -19.38 &    9.81 \\
2015.677 & 2.16547 & 0.00075 &  -87.25 & 4.15 & -66.47 &  -37.28 \\
2016.207 & 2.16450 & 0.00099 & -221.59 & 1.38 &  67.86 &   97.05 \\
2016.285 & 2.16448 & 0.00090 & -224.36 & 1.38 &  70.63 &   99.82 \\
2016.291 & 2.16453 & 0.00094 & -217.44 & 2.76 &  63.70 &   92.89 \\
2016.530 & 2.16522 & 0.00111 & -121.87 & 8.30 & -31.85 &   -2.66 \\
2017.215 & 2.16449 & 0.00114 & -222.98 & 4.15 &  69.24 &   98.42 \\
2017.218 & 2.16450 & 0.00075 & -221.59 & 1.38 &  67.86 &   97.05 \\
2018.226 & 2.16449 & 0.00075 & -222.98 & 4.15 &  69.24 &   98.42 \\
2018.235 & 2.16452 & 0.00077 & -218.82 & 1.38 &  65.09 &   94.28 \\
2018.341 & 2.16453 & 0.00134 & -217.44 & 4.15 &  63.70 &   92.89 \\
2018.422 & 2.16498 & 0.00094 & -155.11 & 1.38 &   1.38 &   30.57 \\
2018.433 & 2.16501 & 0.00130 & -150.96 & 2.76 &  -2.76 &   26.43 \\
2018.505 & 2.16502 & 0.00074 & -149.57 & 1.38 &  -4.15 &   25.04 \\
2019.302 & 2.16450 & 0.00105 & -221.59 & 1.38 &  67.86 &   97.05 \\
2019.336 & 2.16452 & 0.00091 & -218.82 & 2.76 &  65.09 &   94.28 \\
2019.396 & 2.16477 & 0.00107 & -184.20 & 6.92 &  30.46 &   59.65 \\
2019.430 & 2.16499 & 0.00114 & -153.73 & 2.76 &   0.00 &   29.19 \\
2019.438 & 2.16505 & 0.00141 & -145.42 & 1.38 &  -8.30 &   20.89 \\
2022.450 & 2.16446 & 0.00081 & -227.14 & 1.38 &  73.42 &  102.61 \\
2022.485 & 2.16458 & 0.00094 & -210.51 & 1.38 &  56.79 &   85.98 \\
        \hline
    \end{tabular}
       \caption{{\bf Radial velocities measured from the SINFONI and ERIS observations carried out between 2015-2022.} Like Table \ref{tab:rv_values_1} and Table \ref{tab:rv_values_2}, we indicate all relevant observables that are used to derive the best-fit RV model displayed in the periodic pattern of D9. Here, v$_{\rm mod}$ indicates the final RV velocity. The data points from 2022 were observed with ERIS.}
     \label{tab:rv_values_3}
\end{table*}

\newpage

\begin{table*}
        \centering
        \begin{tabular}{|ccccccccc|}
        \hline
          Epoch & Observation ID  & \multicolumn{3}{c}{Data quality} & Exp. Time [s] & \multicolumn{3}{c|}{SNR [a.u.]} \\ 
        \hhline{|~|~|---|~|---|} &  & Total & Medium & Good &  & Noise & Peak & Ratio \\
        \hline    
        2005.463 & 075.B-0547(B) &  21 &   2  &  19  &    60  &  0.32 & 2.22 &  6.93 \\
        2006.210 & 076.B-0259(B) &  5  &   0  &  3   &    600 &  0.84 & 2.28 &  2.71 \\
        2006.221 & 076.B-0259(B) &  2  &   2  &  0   &    600 &  0.54 & 2.05 &  3.79 \\
        2006.627 & 077.B-0503(C) &  5  &   0  &  5   &    600 &  0.39 & 3.58 &  9.20 \\
        2006.710 & 077.B-0503(C) &  3  &   0  &  3   &    600 &  0.35 & 3.39 &  9.68 \\
        2007.235 & 078.B-0520(A) &  8  &   1  &  2   &    600 &  1.32 & 3.12 &  2.36 \\
        2007.355 & 179.B-0261(F) &  10 &   0  &  0   &    600 &  0.89 & 2.84 &  3.19 \\
        2007.557 & 179.B-0261(F) &  3  &   0  &  2   &    600 &  0.23 & 1.60 &  6.95 \\
        2007.562 & 179.B-0261(Z) &  7  &   0  &  7   &    600 &  1.72 & 3.54 &  2.05 \\
        2007.672 & 179.B-0261(K) &  11 &   1  &  5   &    600 &  3.01 & 5.47 &  1.81 \\
        2008.263 & 081.B-0568(A) &  16 &   0  &  15  &    600 &  1.29 & 2.80 &  2.17 \\
        2008.266 & 081.B-0568(A) &  4  &   0  &  4   &    600 &  2.50 & 4.40 &  1.76 \\
        2009.388 & 183.B-0100(B) &  7  &   0  &  7   &    600 &  0.65 & 1.41 &  2.16 \\
        2009.390 & 183.B-0100(B) &  4  &   0  &  4   &    400 &  0.38 & 2.15 &  5.65 \\
        2009.396 & 183.B-0100(B) &  3  &   0  &  3   &    600 &  0.59 & 2.33 &  3.94 \\
        2010.360 & 183.B-0100(O) &  3  &   0  &  3   &    600 &  0.69 & 1.53 &  2.21 \\
        2010.363 & 183.B-0100(O) &  5  &   0  &  5   &    600 &  0.35 & 1.45 &  4.14 \\
        2010.441 & 183.B-0100(O) &  13 &   0  &  13  &    600 &  1.91 & 4.38 &  2.29 \\
        2011.357 & 087.B-0117(I) &  3  &   0  &  3   &    600 &  0.33 & 1.89 &  5.72 \\
        2011.360 & 087.B-0117(I) &  10 &   1  &  9   &    600 &  1.29 & 2.69 &  2.08 \\
        2011.363 & 087.B-0117(I) &  6  &   0  &  6   &    600 &  0.57 & 2.26 &  3.96 \\
        2011.441 & 087.B-0117(I) &  2  &   0  &  2   &    600 &  1.34 & 3.04 &  2.26 \\
        2012.344 & 288.B-5040(A) &  2  &   0  &  2   &    600 &  0.60 & 2.52 &  4.20 \\
        2012.385 & 087.B-0117(J) &  3  &   0  &  3   &    600 &  0.22 & 1.93 &  8.77 \\
        2012.485 & 087.B-0117(J) &  1  &   0  &  1   &    600 &  0.86 & 3.20 &  3.72 \\
        2012.496 & 288.B-5040(A) &  12 &   0  &  10  &    600 &  0.63 & 3.91 &  6.20 \\
        2012.500 & 288.B-5040(A) &  4  &   0  &  4   &    600 &  1.73 & 4.67 &  2.69 \\
        2012.516 & 288.B-5040(A) &  13 &   3  &  8   &    600 &  0.30 & 1.35 &  4.50 \\
        2012.519 & 087.B-0117(J) &  2  &   1  &  1   &    600 &  0.66 & 1.78 &  2.69 \\
        2012.702 & 087.B-0117(J) &  2  &   0  &  2   &    600 &  0.53 & 2.38 &  4.49 \\  
        \hline 
        \end{tabular}   
        \caption{{\bf Analyzed SINFONI data of 2005 to 2012.}}
        \label{tab:data_sinfo1}
        \end{table*}

\newpage        
        
\begin{table*}
        \centering
        \begin{tabular}{|ccccccccc|}
        \hline
          Epoch & Observation ID  & \multicolumn{3}{c}{Data quality} & Exp. Time [s] & \multicolumn{3}{c|}{SNR [a.u.]} \\ 
        \hhline{|~|~|---|~|---|} &  & Total & Medium & Good &  & Noise & Peak & Ratio \\
        \hline 
        2013.260 & 091.B-0088(A)  & 2   &   0  &  2   &    600  & 1.34  & 3.53 & 2.63  \\
        2013.263 & 091.B-0088(A)  & 8   &   0  &  8   &    600  & 1.39  & 2.60 & 1.87  \\
        2013.271 & 091.B-0088(A)  & 16  &   0  & 11   &    600  & 0.79  & 1.80 & 2.27  \\
        2013.274 & 091.B-0088(A)  & 3   &   0  &  3   &    600  & 2.17  & 4.09 & 1.88  \\
        2013.285 & 091.B-0088(A)  & 8   &   1  &  7   &    600  & 0.31  & 1.88 & 6.06  \\
        2013.291 & 091.B-0088(A)  & 3   &   0  &  3   &    600  & 0.30  & 2.23 & 7.43  \\
        2013.296 & 091.B-0088(A)  & 24  &   0  & 24   &    600  & 0.76  & 3.75 & 4.93  \\   
        2013.657 & 091.B-0088(B)  & 10  &   1  &  6   &    600  & 0.25  & 1.56 & 6.24  \\
        2013.662 & 091.B-0088(B)  & 7   &   2  &  4   &    600  & 0.78  & 5.25 & 6.73  \\       
        2013.713 & 091.B-0086(A)  & 6   &   0  &  6   &    600  & 0.10  & 0.53 & 5.30  \\
        2013.729 & 091.B-0086(A)  & 2   &   1  &  0   &    600  & 3.25  & 5.79 & 1.78  \\
        2013.735 & 091.B-0086(A)  & 3   &   1  &  1   &    600  & 0.54  & 2.78 & 5.14  \\   
        2014.154 & 092.B-0920(A)  &	4	&   1  &  3	  &    600	& 0.30 	& 1.42 & 4.73  \\
        2014.157 & 091.B-0183(H)  &	7	&   3  &  1	  &    400	& 0.85 	& 1.72 & 2.02  \\
        2014.166 & 091.B-0183(H)  &	11	&   2  &  4	  &    400	& 1.32 	& 3.06 & 2.31  \\
        2014.177 & 091.B-0183(H)  &	3	&   0  &  3	  &    400	& 0.37 	& 2.00 & 5.40  \\
        2014.180 & 091.B-0183(H)  &	3	&   0  &  3	  &    400	& 1.03 	& 3.10 & 3.00  \\
        2014.183 & 091.B-0183(H)  &	3	&   0  &  3	  &    400	& 0.29 	& 1.05 & 3.62  \\
        2014.185 & 091.B-0183(H)  &	3	&   0  &  3	  &    400	& 0.27 	& 2.08 & 7.70  \\
        2014.194 & 092.B-0920(A)  &	6	&   0  &  6	  &    400	& 0.24 	& 1.16 & 4.83  \\
        2014.196 & 092.B-0920(A)  &	10	&   0  &  10  &    400	& 0.50 	& 5.73 & 2.19 \\
        2014.312 & 092.B-0009(C)  &	13	&   0  &  13  &    400	& 0.49 	& 1.89 & 3.85  \\
        2014.438 & 092.B-0009(C)  &	20	&   0  &  12  &    400	& 0.47 	& 1.27 & 2.70  \\
        2014.519 & 093.B-0092(E)  &	14	&   3  &  0	  &    400	& 0.72 	& 1.65 & 2.29  \\
        2014.651 & 092.B-0398(A)  &	6	&   1  &  3	  &    600	& 1.49 	& 4.14 & 2.77  \\
        2014.665 & 093.B-0092(G)  &	4	&   3  &  0	  &    400	& 0.36 	& 1.05 & 2.91  \\
        2014.683 & 093.B-0218(B)  &	7	&   0  &  7	  &    600	& 1.33  & 3.15 & 2.36  \\  
        \hline
        \end{tabular}
        \caption{{\bf Analyzed SINFONI data of 2013 and 2014.}}
        \label{tab:data_sinfo2}
        \end{table*} 
\newpage        
        
\begin{table*}
        \centering
        \begin{tabular}{|ccccccccc|}
        \hline 
          Epoch & Observation ID  & \multicolumn{3}{c}{Data quality} & Exp. Time [s] & \multicolumn{3}{c|}{SNR [a.u.]} \\ 
        \hhline{|~|~|---|~|---|} &  & Total & Medium & Good &  & Noise & Peak & Ratio \\
        \hline 
        2015.302 & 095.B-0036(A)  & 12 &   0  &  12 & 600 & 0.47 & 2.43 & 5.17 \\
        2015.307 & 095.B-0036(A)  & 9  &   0  &  9  & 600 & 0.38 & 2.18 & 5.73 \\
        2015.323 & 095.B-0036(A)  & 14 &   0  &  14 & 600 & 2.66 & 6.16 & 2.31 \\
        2015.326 & 095.B-0036(A)  & 12 &   0  &  8  & 600 & 0.97 & 4.11 & 4.23 \\
        2015.379 & 095.B-0036(A)  & 3  &   0  &  3  & 600 & 1.11 & 5.39 & 4.85 \\
        2015.583 & 095.B-0036(C)  & 23 &   7  &  8  & 400 & 0.78 & 3.84 & 4.92 \\
        2015.677 & 095.B-0036(D)  & 17 &  11  &  4  & 400 & 0.18 & 1.49 & 8.27 \\
        2016.207 & 096.B-0157(B)  & 17 &   0  &  17 & 400 & 0.66 & 1.35 & 2.04 \\
        2016.285 & 594.B-0498(R)  & 12 &   0  &  12 & 600 & 0.24 & 1.67 & 6.95 \\
        2016.291 & 594.B-0498(R)  & 10 &   0  &  8  & 600 & 1.57 & 3.36 & 2.14 \\
        2016.530 & 097.B-0050(A)  & 27 &   0  &  13 & 600 & 0.06 & 0.39 & 6.50 \\        
        2017.215 & 598.B-0043(D)  & 11 &   0  &  5  & 600 & 0.16 & 1.59 & 9.93 \\
        2017.218 & 598.B-0043(D)  & 15 &   4  &  11 & 600 & 0.15 & 1.33 & 8.86 \\     
        2018.226 & 598.B-0043(D)  &  8 &   0  &  8  & 600 & 0.12 & 0.97 & 8.08 \\
        2018.235 & 598.B-0043(D)  & 12 &   1  &  9  & 600 & 0.24 & 1.18 & 4.91 \\
        2018.341 & 598.B-0043(E)  & 17 &   0  &  17 & 600 & 0.17 & 1.61 & 9.47 \\
        2018.422 & 598.B-0043(F)  &  8 &   0  &  8  & 600 & 1.61 & 5.07 & 3.14 \\
        2018.433 & 598.B-0043(F)  & 14 &   1  &  7  & 600 & 0.25 & 1.18 & 4.72 \\
        2018.505 & 598.B-0043(G)  & 22 &  12  &  10 & 600 & 0.45 & 1.72 & 3.82 \\
        2019.302 & 0103.B-0026(B) &  9 &   0  &  8  & 600 & 0.10 & 0.96 & 9.60 \\
        2019.336 & 0103.B-0026(F) &  8 &   0  &  8  & 600 & 0.52 & 1.59 & 3.05 \\
        2019.396 & 5102.B-0086(Q) &  4 &   0  &  2  & 600 & 0.89 & 2.37 & 2.66 \\
        2019.430 & 594.B-0498(Q)  & 11 &   0  &  10 & 600 & 0.24 & 1.84 & 7.66 \\
        2019.438 & 5102.B-0086(Q) & 14 &   0  &  10 & 600 & 0.34 & 1.25 & 3.67 \\
        2022.450 & 60.A-9917(C)   & 3  &   0  &  2  & 600 & 20.1 & 104.56  & 5.20 \\
        2022.485 & 60.A-9917(D)   & 3  &   0  &  2  & 600 & 57.75 & 322.58 & 5.58 \\       
        \hline 
        \end{tabular}
        \caption{{\bf Analyzed SINFONI and ERIS data of 2015 to 2022.} The two observations in 2022 were carried out with ERIS.}
        \label{tab:data_sinfo4}
        \end{table*}        
\newpage


\clearpage

\begin{thebibliography}{100}
\expandafter\ifx\csname url\endcsname\relax
  \def\url#1{\burl{#1}}\fi
\expandafter\ifx\csname urlprefix\endcsname\relax\def\urlprefix{URL }\fi
\providecommand{\bibinfo}[2]{#2}
\providecommand{\eprint}[2][]{\url{#2}}
\providecommand{\doi}[1]{\url{https://doi.org/#1}}
\bibcommenthead

\bibitem{Schoedel2009}
\bibinfo{author}{{Sch{\"o}del}, R.}, \bibinfo{author}{{Merritt}, D.} \& \bibinfo{author}{{Eckart}, A.}
\newblock \bibinfo{title}{{The nuclear star cluster of the Milky Way: proper motions and mass}}.
\newblock \emph{\bibinfo{journal}{Astron. Astrophys.}} {\bibinfo{volume}{502}}, \bibinfo{pages}{91--111} (\bibinfo{year}{2009}).

\bibitem{Schoedel2010}
\bibinfo{author}{{Sch{\"o}del}, R.}, \bibinfo{author}{{Najarro}, F.}, \bibinfo{author}{{Muzic}, K.} \& \bibinfo{author}{{Eckart}, A.}
\newblock \bibinfo{title}{{Peering through the veil: near-infrared photometry and extinction for the Galactic nuclear star cluster. Accurate near infrared H, Ks, and L' photometry and the near-infrared extinction-law toward the central parsec of the Galaxy}}.
\newblock \emph{\bibinfo{journal}{Astron. Astrophys.}} {\bibinfo{volume}{511}}, \bibinfo{pages}{A18} (\bibinfo{year}{2010}).

\bibitem{Ali2020}
\bibinfo{author}{{Ali}, B.} \emph{et~al.}
\newblock \bibinfo{title}{{Kinematic Structure of the Galactic Center S Cluster}}.
\newblock \emph{\bibinfo{journal}{Astrophys. J.}} {\bibinfo{volume}{896}}, \bibinfo{pages}{100} (\bibinfo{year}{2020}).

\bibitem{Eckart1996Natur}
\bibinfo{author}{{Eckart}, A.} \& \bibinfo{author}{{Genzel}, R.}
\newblock \bibinfo{title}{{Observations of stellar proper motions near the Galactic Centre}}.
\newblock \emph{\bibinfo{journal}{Nature}} {\bibinfo{volume}{383}}, \bibinfo{pages}{415--417} (\bibinfo{year}{1996}).

\bibitem{Ghez1998}
\bibinfo{author}{{Ghez}, A.~M.}, \bibinfo{author}{{Klein}, B.~L.}, \bibinfo{author}{{Morris}, M.} \& \bibinfo{author}{{Becklin}, E.~E.}
\newblock \bibinfo{title}{{High Proper-Motion Stars in the Vicinity of Sagittarius A*: Evidence for a Supermassive Black Hole at the Center of Our Galaxy}}.
\newblock \emph{\bibinfo{journal}{apj}} {\bibinfo{volume}{509}}, \bibinfo{pages}{678--686} (\bibinfo{year}{1998}).

\bibitem{Morris1993}
\bibinfo{author}{{Morris}, M.}
\newblock \bibinfo{title}{{Massive Star Formation near the Galactic Center and the Fate of the Stellar Remnants}}.
\newblock \emph{\bibinfo{journal}{Astrophys. J.}} {\bibinfo{volume}{408}}, \bibinfo{pages}{496} (\bibinfo{year}{1993}).

\bibitem{Habibi2019}
\bibinfo{author}{{Habibi}, M.} \emph{et~al.}
\newblock \bibinfo{title}{{Spectroscopic Detection of a Cusp of Late-type Stars around the Central Black Hole in the Milky Way}}.
\newblock \emph{\bibinfo{journal}{Astrophys. J.l}} {\bibinfo{volume}{872}}, \bibinfo{pages}{L15} (\bibinfo{year}{2019}).

\bibitem{Lu2013}
\bibinfo{author}{{Lu}, J.~R.} \emph{et~al.}
\newblock \bibinfo{title}{{Stellar Populations in the Central 0.5 pc of the Galaxy. II. The Initial Mass Function}}.
\newblock \emph{\bibinfo{journal}{Astrophys. J.}} {\bibinfo{volume}{764}}, \bibinfo{pages}{155} (\bibinfo{year}{2013}).

\bibitem{Habibi2017}
\bibinfo{author}{{Habibi}, M.} \emph{et~al.}
\newblock \bibinfo{title}{{Twelve Years of Spectroscopic Monitoring in the Galactic Center: The Closest Look at S-stars near the Black Hole}}.
\newblock \emph{\bibinfo{journal}{Astrophys. J.}} {\bibinfo{volume}{847}}, \bibinfo{pages}{120} (\bibinfo{year}{2017}).

\bibitem{Ghez2003}
\bibinfo{author}{{Ghez}, A.~M.} \emph{et~al.}
\newblock \bibinfo{title}{{The First Measurement of Spectral Lines in a Short-Period Star Bound to the Galaxy's Central Black Hole: A Paradox of Youth}}.
\newblock \emph{\bibinfo{journal}{apjl}} {\bibinfo{volume}{586}}, \bibinfo{pages}{L127--L131} (\bibinfo{year}{2003}).

\bibitem{Chu2023}
\bibinfo{author}{{Chu}, D.~S.} \emph{et~al.}
\newblock \bibinfo{title}{{Evidence of a Decreased Binary Fraction for Massive Stars within 20 milliparsecs of the Supermassive Black Hole at the Galactic Center}}.
\newblock \emph{\bibinfo{journal}{Astrophys. J.}} {\bibinfo{volume}{948}}, \bibinfo{pages}{94} (\bibinfo{year}{2023}).

\bibitem{Offner2023}
\bibinfo{author}{{Offner}, S.~S.~R.} \emph{et~al.}
\newblock \bibinfo{editor}{{Inutsuka}, S.}, \bibinfo{editor}{{Aikawa}, Y.}, \bibinfo{editor}{{Muto}, T.}, \bibinfo{editor}{{Tomida}, K.} \& \bibinfo{editor}{{Tamura}, M.} (eds) \emph{\bibinfo{title}{{The Origin and Evolution of Multiple Star Systems}}}.
\newblock (eds \bibinfo{editor}{{Inutsuka}, S.}, \bibinfo{editor}{{Aikawa}, Y.}, \bibinfo{editor}{{Muto}, T.}, \bibinfo{editor}{{Tomida}, K.} \& \bibinfo{editor}{{Tamura}, M.}) \emph{\bibinfo{booktitle}{Protostars and Planets VII}}, Vol. \bibinfo{volume}{534} of \emph{\bibinfo{series}{Astronomical Society of the Pacific Conference Series}}, \bibinfo{pages}{275} (\bibinfo{year}{2023}).
\newblock \eprint{2203.10066}.

\bibitem{Michaely2023}
\bibinfo{author}{{Michaely}, E.} \& \bibinfo{author}{{Naoz}, S.}
\newblock \bibinfo{title}{{New Dynamical Channel: Wide Binaries in the Galactic Center as a Source of Binary Interactions}}.
\newblock \emph{\bibinfo{journal}{arXiv e-prints}} \bibinfo{pages}{arXiv:2310.02558} (\bibinfo{year}{2023}).

\bibitem{Gautam2024}
\bibinfo{author}{{Gautam}, A.~K.} \emph{et~al.}
\newblock \bibinfo{title}{{An Estimate of the Binary Star Fraction Among Young Stars at the Galactic Center: Possible Evidence of a Radial Dependence}}.
\newblock \emph{\bibinfo{journal}{arXiv e-prints}} \bibinfo{pages}{arXiv:2401.12555} (\bibinfo{year}{2024}).

\bibitem{Sana2012}
\bibinfo{author}{{Sana}, H.} \emph{et~al.}
\newblock \bibinfo{title}{{Binary Interaction Dominates the Evolution of Massive Stars}}.
\newblock \emph{\bibinfo{journal}{Science}} {\bibinfo{volume}{337}}, \bibinfo{pages}{444} (\bibinfo{year}{2012}).

\bibitem{Ciurlo2020}
\bibinfo{author}{{Ciurlo}, A.} \emph{et~al.}
\newblock \bibinfo{title}{{A population of dust-enshrouded objects orbiting the Galactic black hole}}.
\newblock \emph{\bibinfo{journal}{nat}} {\bibinfo{volume}{577}}, \bibinfo{pages}{337--340} (\bibinfo{year}{2020}).

\bibitem{Peissker2024}
\bibinfo{author}{{Pei{\ss}ker}, F.} \emph{et~al.}
\newblock \bibinfo{title}{{Candidate young stellar objects in the S-cluster: Kinematic analysis of a subpopulation of the low-mass G objects close to Sgr A*}}.
\newblock \emph{\bibinfo{journal}{Astron. Astrophys.}} {\bibinfo{volume}{686}}, \bibinfo{pages}{A235} (\bibinfo{year}{2024}).

\bibitem{Eisenhauer2003}
\bibinfo{author}{{Eisenhauer}, F.} \emph{et~al.}
\newblock \bibinfo{editor}{{Iye}, M.} \& \bibinfo{editor}{{Moorwood}, A.~F.~M.} (eds) \emph{\bibinfo{title}{{SINFONI - Integral field spectroscopy at 50 milli-arcsecond resolution with the ESO VLT}}}.
\newblock (eds \bibinfo{editor}{{Iye}, M.} \& \bibinfo{editor}{{Moorwood}, A.~F.~M.}) \emph{\bibinfo{booktitle}{Instrument Design and Performance for Optical/Infrared Ground-based Telescopes}}, Vol. \bibinfo{volume}{4841} of \emph{\bibinfo{series}{Proc. SPIE}}, \bibinfo{pages}{1548--1561} (\bibinfo{year}{2003}).
\newblock \eprint{astro-ph/0306191}.


\bibitem{Bonnet2004}
\bibinfo{author}{{Bonnet}, H.} \emph{et~al.}
\newblock \bibinfo{title}{{First light of SINFONI at the VLT}}.
\newblock \emph{\bibinfo{journal}{The Messenger}} {\bibinfo{volume}{117}}, \bibinfo{pages}{17--24} (\bibinfo{year}{2004}).

\bibitem{Peissker2020b}
\bibinfo{author}{{Pei{\ss}ker}, F.} \emph{et~al.}
\newblock \bibinfo{title}{{Monitoring dusty sources in the vicinity of Sagittarius A*}}.
\newblock \emph{\bibinfo{journal}{Astron. Astrophys.}} {\bibinfo{volume}{634}}, \bibinfo{pages}{A35} (\bibinfo{year}{2020b}).

\bibitem{Davies2023}
\bibinfo{author}{{Davies}, R.} \emph{et~al.}
\newblock \bibinfo{title}{{The Enhanced Resolution Imager and Spectrograph for the VLT}}.
\newblock \emph{\bibinfo{journal}{Astron. Astrophys.}} {\bibinfo{volume}{674}}, \bibinfo{pages}{A207} (\bibinfo{year}{2023}).

\bibitem{Peissker2022}
\bibinfo{author}{{Pei{\ss}ker}, F.}, \bibinfo{author}{{Eckart}, A.}, \bibinfo{author}{{Zaja{\v{c}}ek}, M.} \& \bibinfo{author}{{Britzen}, S.}
\newblock \bibinfo{title}{{Observation of S4716-a Star with a 4 yr Orbit around Sgr A*}}.
\newblock \emph{\bibinfo{journal}{Astrophys. J.}} {\bibinfo{volume}{933}}, \bibinfo{pages}{49} (\bibinfo{year}{2022}).

\bibitem{eht2022}
\bibinfo{author}{{Event Horizon Telescope Collaboration}} \emph{et~al.}
\newblock \bibinfo{title}{{First Sagittarius A* Event Horizon Telescope Results. I. The Shadow of the Supermassive Black Hole in the Center of the Milky Way}}.
\newblock \emph{\bibinfo{journal}{Astrophys. J.l}} {\bibinfo{volume}{930}}, \bibinfo{pages}{L12} (\bibinfo{year}{2022}).

\bibitem{Gillessen2017}
\bibinfo{author}{{Gillessen}, S.} \emph{et~al.}
\newblock \bibinfo{title}{{An Update on Monitoring Stellar Orbits in the Galactic Center}}.
\newblock \emph{\bibinfo{journal}{Astrophys. J.}} {\bibinfo{volume}{837}}, \bibinfo{pages}{30} (\bibinfo{year}{2017}).

\bibitem{Matthews1994}
\bibinfo{author}{{Matthews}, K.} \& \bibinfo{author}{{Soifer}, B.~T.}
\newblock \bibinfo{editor}{{McLean}, I.~S.} (ed.) \emph{\bibinfo{title}{{The Near Infrared Camera on the W. M. Keck Telescope}}}.
\newblock (ed.\bibinfo{editor}{{McLean}, I.~S.}) \emph{\bibinfo{booktitle}{Astronomy with Arrays, The Next Generation}}, Vol. \bibinfo{volume}{190} of \emph{\bibinfo{series}{Astrophysics and Space Science Library}}, \bibinfo{pages}{239} (\bibinfo{year}{1994}).

\bibitem{Tran2016}
\bibinfo{author}{{Tran}, H.~D.} \emph{et~al.}
\newblock \bibinfo{editor}{{Peck}, A.~B.}, \bibinfo{editor}{{Seaman}, R.~L.} \& \bibinfo{editor}{{Benn}, C.~R.} (eds) \emph{\bibinfo{title}{{Data reduction pipelines for the Keck Observatory Archive}}}.
\newblock (eds \bibinfo{editor}{{Peck}, A.~B.}, \bibinfo{editor}{{Seaman}, R.~L.} \& \bibinfo{editor}{{Benn}, C.~R.}) \emph{\bibinfo{booktitle}{Observatory Operations: Strategies, Processes, and Systems VI}}, Vol. \bibinfo{volume}{9910} of \emph{\bibinfo{series}{Society of Photo-Optical Instrumentation Engineers (SPIE) Conference Series}}, \bibinfo{pages}{99102E} (\bibinfo{year}{2016}).

\bibitem{Eckart2013}
\bibinfo{author}{{Eckart}, A.} \emph{et~al.}
\newblock \bibinfo{title}{Near-infrared proper motions and spectroscopy of infrared excess sources at the galactic center}.
\newblock \emph{\bibinfo{journal}{Astron. Astrophys.}} {\bibinfo{volume}{551}}, \bibinfo{pages}{A18} (\bibinfo{year}{2013}).

\bibitem{Witzel2017}
\bibinfo{author}{{Witzel}, G.} \emph{et~al.}
\newblock \bibinfo{title}{{The Post-periapsis Evolution of Galactic Center Source G1: The Second Case of a Resolved Tidal Interaction with a Supermassive Black Hole}}.
\newblock \emph{\bibinfo{journal}{Astrophys. J.}} {\bibinfo{volume}{847}}, \bibinfo{pages}{80} (\bibinfo{year}{2017}).

\bibitem{peissker2021c}
\bibinfo{author}{{Pei{\ss}ker}, F.} \emph{et~al.}
\newblock \bibinfo{title}{{The Apparent Tail of the Galactic Center Object G2/DSO}}.
\newblock \emph{\bibinfo{journal}{Astrophys. J.}} {\bibinfo{volume}{923}}, \bibinfo{pages}{69} (\bibinfo{year}{2021c}).

\bibitem{Scoville2013}
\bibinfo{author}{{Scoville}, N.} \& \bibinfo{author}{{Burkert}, A.}
\newblock \bibinfo{title}{The galactic center cloud g2, a young low-mass star with a stellar wind}.
\newblock \emph{\bibinfo{journal}{Astrophys. J.}} {\bibinfo{volume}{768}}, \bibinfo{pages}{108} (\bibinfo{year}{2013}).

\bibitem{Valencia-S.2015}
\bibinfo{author}{{Valencia-S.}, M.} \emph{et~al.}
\newblock \bibinfo{title}{{Monitoring the Dusty S-cluster Object (DSO/G2) on its Orbit toward the Galactic Center Black Hole}}.
\newblock \emph{\bibinfo{journal}{Astrophys. J.}} {\bibinfo{volume}{800}}, \bibinfo{pages}{125} (\bibinfo{year}{2015}).

\bibitem{Robitaille2017}
\bibinfo{author}{{Robitaille}, T.~P.}
\newblock \bibinfo{title}{{A modular set of synthetic spectral energy distributions for young stellar objects}}.
\newblock \emph{\bibinfo{journal}{Astron. Astrophys.}} {\bibinfo{volume}{600}}, \bibinfo{pages}{A11} (\bibinfo{year}{2017}).

\bibitem{Leggett2003}
\bibinfo{author}{{Leggett}, S.~K.} \emph{et~al.}
\newblock \bibinfo{title}{{L' and M' standard stars for the Mauna Kea Observatories Near-Infrared system}}.
\newblock \emph{\bibinfo{journal}{Mon. Not. R. Soc.}} {\bibinfo{volume}{345}}, \bibinfo{pages}{144--152} (\bibinfo{year}{2003}).

\bibitem{Eckart2002}
\bibinfo{author}{{Eckart}, A.}, \bibinfo{author}{{Genzel}, R.}, \bibinfo{author}{{Ott}, T.} \& \bibinfo{author}{{Sch{\"o}del}, R.}
\newblock \bibinfo{title}{{Stellar orbits near Sagittarius A*}}.
\newblock \emph{\bibinfo{journal}{Mon. Not. R. Soc.}} {\bibinfo{volume}{331}}, \bibinfo{pages}{917--934} (\bibinfo{year}{2002}).

\bibitem{exostriker2019}
\bibinfo{author}{{Trifonov}, T.}
\newblock \bibinfo{title}{{The Exo-Striker: Transit and radial velocity interactive fitting tool for orbital analysis and N-body simulations}}.
\newblock \bibinfo{howpublished}{Astrophysics Source Code Library, record ascl:1906.004} (\bibinfo{year}{2019}).
\newblock \eprint{1906.004}.

\bibitem{Bally2005}
\bibinfo{author}{{Bally}, J.} \& \bibinfo{author}{{Zinnecker}, H.}
\newblock \bibinfo{title}{{The Birth of High-Mass Stars: Accretion and/or Mergers?}}
\newblock \emph{\bibinfo{journal}{Astron. J.}} {\bibinfo{volume}{129}}, \bibinfo{pages}{2281--2293} (\bibinfo{year}{2005}).

\bibitem{Baines2006}
\bibinfo{author}{{Baines}, D.}, \bibinfo{author}{{Oudmaijer}, R.~D.}, \bibinfo{author}{{Porter}, J.~M.} \& \bibinfo{author}{{Pozzo}, M.}
\newblock \bibinfo{title}{{On the binarity of Herbig Ae/Be stars}}.
\newblock \emph{\bibinfo{journal}{Mon. Not. R. Soc.}} {\bibinfo{volume}{367}}, \bibinfo{pages}{737--753} (\bibinfo{year}{2006}).

\bibitem{Tanaka2016}
\bibinfo{author}{{Tanaka}, K. E.~I.}, \bibinfo{author}{{Tan}, J.~C.} \& \bibinfo{author}{{Zhang}, Y.}
\newblock \bibinfo{title}{{Outflow-confined HII Regions. I. First Signposts of Massive Star Formation}}.
\newblock \emph{\bibinfo{journal}{Astrophys. J.}} {\bibinfo{volume}{818}}, \bibinfo{pages}{52} (\bibinfo{year}{2016}).

\bibitem{Tambovtseva2016}
\bibinfo{author}{{Tambovtseva}, L.~V.}, \bibinfo{author}{{Grinin}, V.~P.} \& \bibinfo{author}{{Weigelt}, G.}
\newblock \bibinfo{title}{{Brackett {\ensuremath{\gamma}} radiation from the inner gaseous accretion disk, magnetosphere, and disk wind region of Herbig AeBe stars}}.
\newblock \emph{\bibinfo{journal}{Astron. Astrophys.}} {\bibinfo{volume}{590}}, \bibinfo{pages}{A97} (\bibinfo{year}{2016}).

\bibitem{Fateeva2011}
\bibinfo{author}{{Fateeva}, A.~M.}, \bibinfo{author}{{Bisikalo}, D.~V.}, \bibinfo{author}{{Kaygorodov}, P.~V.} \& \bibinfo{author}{{Sytov}, A.~Y.}
\newblock \bibinfo{title}{{Gaseous flows in the inner part of the circumbinary disk of the T Tauri star}}.
\newblock \emph{\bibinfo{journal}{Astrophys. Space Sci.}} {\bibinfo{volume}{335}}, \bibinfo{pages}{125--129} (\bibinfo{year}{2011}).

\bibitem{Muzerolle1998}
\bibinfo{author}{{Muzerolle}, J.}, \bibinfo{author}{{Hartmann}, L.} \& \bibinfo{author}{{Calvet}, N.}
\newblock \bibinfo{title}{{A Brgamma Probe of Disk Accretion in T Tauri Stars and Embedded Young Stellar Objects}}.
\newblock \emph{\bibinfo{journal}{Astron. J.}} {\bibinfo{volume}{116}}, \bibinfo{pages}{2965--2974} (\bibinfo{year}{1998}).

\bibitem{Grant2022}
\bibinfo{author}{{Grant}, S.~L.}, \bibinfo{author}{{Espaillat}, C.~C.}, \bibinfo{author}{{Brittain}, S.}, \bibinfo{author}{{Scott-Joseph}, C.} \& \bibinfo{author}{{Calvet}, N.}
\newblock \bibinfo{title}{{Tracing Accretion onto Herbig Ae/Be Stars Using the Br{\ensuremath{\gamma}} Line}}.
\newblock \emph{\bibinfo{journal}{Astrophys. J.}} {\bibinfo{volume}{926}}, \bibinfo{pages}{229} (\bibinfo{year}{2022}).

\bibitem{Fiorellino2022}
\bibinfo{author}{{Fiorellino}, E.}, \bibinfo{author}{{Park}, S.}, \bibinfo{author}{{K{\'o}sp{\'a}l}, {\'A}.} \& \bibinfo{author}{{{\'A}brah{\'a}m}, P.}
\newblock \bibinfo{title}{{The Accretion Process in the DQ Tau Binary System}}.
\newblock \emph{\bibinfo{journal}{Astrophys. J.}} {\bibinfo{volume}{928}}, \bibinfo{pages}{81} (\bibinfo{year}{2022}).

\bibitem{Friedjung2010}
\bibinfo{author}{{Friedjung}, M.}, \bibinfo{author}{{Miko{\l}ajewska}, J.}, \bibinfo{author}{{Zajczyk}, A.} \& \bibinfo{author}{{Eriksson}, M.}
\newblock \bibinfo{title}{{UV emission line shifts of symbiotic binaries}}.
\newblock \emph{\bibinfo{journal}{Astron. Astrophys.}} {\bibinfo{volume}{512}}, \bibinfo{pages}{A80} (\bibinfo{year}{2010}).

\bibitem{Horne1986}
\bibinfo{author}{{Horne}, K.} \& \bibinfo{author}{{Marsh}, T.~R.}
\newblock \bibinfo{title}{{Emission line formation in accretion discs}}.
\newblock \emph{\bibinfo{journal}{Mon. Not. R. Soc.}} {\bibinfo{volume}{218}}, \bibinfo{pages}{761--773} (\bibinfo{year}{1986}).

\bibitem{Kraus2012}
\bibinfo{author}{{Kraus}, S.} \emph{et~al.}
\newblock \bibinfo{title}{{Gas Distribution, Kinematics, and Excitation Structure in the Disks around the Classical Be Stars {\ensuremath{\beta}} Canis Minoris and {\ensuremath{\zeta}} Tauri}}.
\newblock \emph{\bibinfo{journal}{Astrophys. J.}} {\bibinfo{volume}{744}}, \bibinfo{pages}{19} (\bibinfo{year}{2012}).

\bibitem{Kraus2009}
\bibinfo{author}{{Kraus}, S.} \emph{et~al.}
\newblock \bibinfo{title}{{Tracing the young massive high-eccentricity binary system {\ensuremath{\theta}}\^1Orionis C through periastron passage}}.
\newblock \emph{\bibinfo{journal}{Astron. Astrophys.}} {\bibinfo{volume}{497}}, \bibinfo{pages}{195--207} (\bibinfo{year}{2009}).

\bibitem{Frost2022}
\bibinfo{author}{{Frost}, A.~J.} \emph{et~al.}
\newblock \bibinfo{title}{{HR 6819 is a binary system with no black hole. Revisiting the source with infrared interferometry and optical integral field spectroscopy}}.
\newblock \emph{\bibinfo{journal}{Astron. Astrophys.}} {\bibinfo{volume}{659}}, \bibinfo{pages}{L3} (\bibinfo{year}{2022}).

\bibitem{Garcia2013}
\bibinfo{author}{{Garcia}, P.~J.~V.} \emph{et~al.}
\newblock \bibinfo{title}{{Pre-main-sequence binaries with tidally disrupted discs: the Br{\ensuremath{\gamma}} in HD 104237}}.
\newblock \emph{\bibinfo{journal}{Mon. Not. R. Soc.}} {\bibinfo{volume}{430}}, \bibinfo{pages}{1839--1853} (\bibinfo{year}{2013}).

\bibitem{Hartmann1998}
\bibinfo{author}{{Hartmann}, L.}, \bibinfo{author}{{Calvet}, N.}, \bibinfo{author}{{Gullbring}, E.} \& \bibinfo{author}{{D'Alessio}, P.}
\newblock \bibinfo{title}{{Accretion and the Evolution of T Tauri Disks}}.
\newblock \emph{\bibinfo{journal}{Astrophys. J.}} {\bibinfo{volume}{495}}, \bibinfo{pages}{385--400} (\bibinfo{year}{1998}).

\bibitem{Grinin2010}
\bibinfo{author}{{Grinin}, V.~P.}, \bibinfo{author}{{Rostopchina}, A.~N.}, \bibinfo{author}{{Barsunova}, O.~Y.} \& \bibinfo{author}{{Demidova}, T.~V.}
\newblock \bibinfo{title}{{Mechanism for cyclical activity of the Herbig Ae star BF Ori}}.
\newblock \emph{\bibinfo{journal}{Astrophysics}} {\bibinfo{volume}{53}}, \bibinfo{pages}{367--372} (\bibinfo{year}{2010}).

\bibitem{Wheelwright2010}
\bibinfo{author}{{Wheelwright}, H.~E.}, \bibinfo{author}{{Oudmaijer}, R.~D.} \& \bibinfo{author}{{Goodwin}, S.~P.}
\newblock \bibinfo{title}{{The mass ratio and formation mechanisms of Herbig Ae/Be star binary systems}}.
\newblock \emph{\bibinfo{journal}{Mon. Not. R. Soc.}} {\bibinfo{volume}{401}}, \bibinfo{pages}{1199--1218} (\bibinfo{year}{2010}).

\bibitem{Lutzgendorf2016}
\bibinfo{author}{{L{\"u}tzgendorf}, N.}, \bibinfo{author}{{Helm}, E. v.~d.}, \bibinfo{author}{{Pelupessy}, F.~I.} \& \bibinfo{author}{{Portegies Zwart}, S.}
\newblock \bibinfo{title}{{Stellar winds near massive black holes - the case of the S-stars}}.
\newblock \emph{\bibinfo{journal}{Mon. Not. R. Soc.}} {\bibinfo{volume}{456}}, \bibinfo{pages}{3645--3654} (\bibinfo{year}{2016}).

\bibitem{Hollenbach1994}
\bibinfo{author}{{Hollenbach}, D.}, \bibinfo{author}{{Johnstone}, D.}, \bibinfo{author}{{Lizano}, S.} \& \bibinfo{author}{{Shu}, F.}
\newblock \bibinfo{title}{{Photoevaporation of Disks around Massive Stars and Application to Ultracompact H II Regions}}.
\newblock \emph{\bibinfo{journal}{Astrophys. J.}} {\bibinfo{volume}{428}}, \bibinfo{pages}{654} (\bibinfo{year}{1994}).

\bibitem{Johnstone1998}
\bibinfo{author}{{Johnstone}, D.}, \bibinfo{author}{{Hollenbach}, D.} \& \bibinfo{author}{{Bally}, J.}
\newblock \bibinfo{title}{{Photoevaporation of Disks and Clumps by Nearby Massive Stars: Application to Disk Destruction in the Orion Nebula}}.
\newblock \emph{\bibinfo{journal}{Astrophys. J.}} {\bibinfo{volume}{499}}, \bibinfo{pages}{758--776} (\bibinfo{year}{1998}).

\bibitem{Messina2019}
\bibinfo{author}{{Messina}, S.}
\newblock \bibinfo{title}{{Evidence from stellar rotation for early disc dispersal owing to close companions}}.
\newblock \emph{\bibinfo{journal}{Astron. Astrophys.}} {\bibinfo{volume}{627}}, \bibinfo{pages}{A97} (\bibinfo{year}{2019}).

\bibitem{Alves2019}
\bibinfo{author}{{Alves}, F.~O.} \emph{et~al.}
\newblock \bibinfo{title}{{Gas flow and accretion via spiral streamers and circumstellar disks in a young binary protostar}}.
\newblock \emph{\bibinfo{journal}{Science}} {\bibinfo{volume}{366}}, \bibinfo{pages}{90--93} (\bibinfo{year}{2019}).

\bibitem{Bressan2012}
\bibinfo{author}{{Bressan}, A.} \emph{et~al.}
\newblock \bibinfo{title}{{PARSEC: stellar tracks and isochrones with the PAdova and TRieste Stellar Evolution Code}}.
\newblock \emph{\bibinfo{journal}{Mon. Not. R. Soc.}} {\bibinfo{volume}{427}}, \bibinfo{pages}{127--145} (\bibinfo{year}{2012}).

\bibitem{Fragione2018}
\bibinfo{author}{{Fragione}, G.} \& \bibinfo{author}{{Gualandris}, A.}
\newblock \bibinfo{title}{{Tidal breakup of triple stars in the Galactic Centre}}.
\newblock \emph{\bibinfo{journal}{Mon. Not. R. Soc.}} {\bibinfo{volume}{475}}, \bibinfo{pages}{4986--4993} (\bibinfo{year}{2018}).

\bibitem{Bonnell2008}
\bibinfo{author}{{Bonnell}, I.~A.} \& \bibinfo{author}{{Rice}, W.~K.~M.}
\newblock \bibinfo{title}{{Star Formation Around Supermassive Black Holes}}.
\newblock \emph{\bibinfo{journal}{Science}} {\bibinfo{volume}{321}}, \bibinfo{pages}{1060} (\bibinfo{year}{2008}).

\bibitem{Jalali2014}
\bibinfo{author}{{Jalali}, B.} \emph{et~al.}
\newblock \bibinfo{title}{{Star formation in the vicinity of nuclear black holes: young stellar objects close to Sgr A*}}.
\newblock \emph{\bibinfo{journal}{Mon. Not. R. Soc.}} {\bibinfo{volume}{444}}, \bibinfo{pages}{1205--1220} (\bibinfo{year}{2014}).

\bibitem{Zajacek2017}
\bibinfo{author}{{Zaja{\v c}ek}, M.} \emph{et~al.}
\newblock \bibinfo{title}{{Nature of the Galactic centre NIR-excess sources. I. What can we learn from the continuum observations of the DSO/G2 source?}}
\newblock \emph{\bibinfo{journal}{Astron. Astrophys.}} {\bibinfo{volume}{602}}, \bibinfo{pages}{A121} (\bibinfo{year}{2017}).

\bibitem{Hills1988}
\bibinfo{author}{{Hills}, J.~G.}
\newblock \bibinfo{title}{{Hyper-velocity and tidal stars from binaries disrupted by a massive Galactic black hole}}.
\newblock \emph{\bibinfo{journal}{Nature}} {\bibinfo{volume}{331}}, \bibinfo{pages}{687--689} (\bibinfo{year}{1988}).

\bibitem{Yu2003}
\bibinfo{author}{{Yu}, Q.} \& \bibinfo{author}{{Tremaine}, S.}
\newblock \bibinfo{title}{{Ejection of Hypervelocity Stars by the (Binary) Black Hole in the Galactic Center}}.
\newblock \emph{\bibinfo{journal}{Astrophys. J.}} {\bibinfo{volume}{599}}, \bibinfo{pages}{1129--1138} (\bibinfo{year}{2003}).

\bibitem{Zajacek2014}
\bibinfo{author}{{Zaja{\v{c}}ek}, M.}, \bibinfo{author}{{Karas}, V.} \& \bibinfo{author}{{Eckart}, A.}
\newblock \bibinfo{title}{{Dust-enshrouded star near supermassive black hole: predictions for high-eccentricity passages near low-luminosity galactic nuclei}}.
\newblock \emph{\bibinfo{journal}{Astron. Astrophys.}} {\bibinfo{volume}{565}}, \bibinfo{pages}{A17} (\bibinfo{year}{2014}).

\bibitem{Armitage2003}
\bibinfo{author}{{Armitage}, P.~J.}, \bibinfo{author}{{Clarke}, C.~J.} \& \bibinfo{author}{{Palla}, F.}
\newblock \bibinfo{title}{{Dispersion in the lifetime and accretion rate of T Tauri discs}}.
\newblock \emph{\bibinfo{journal}{Mon. Not. R. Soc.}} {\bibinfo{volume}{342}}, \bibinfo{pages}{1139--1146} (\bibinfo{year}{2003}).

\bibitem{Alexander2012}
\bibinfo{author}{{Alexander}, R.}
\newblock \bibinfo{title}{{The Dispersal of Protoplanetary Disks around Binary Stars}}.
\newblock \emph{\bibinfo{journal}{Astrophys. J.l}} {\bibinfo{volume}{757}}, \bibinfo{pages}{L29} (\bibinfo{year}{2012}).

\bibitem{Zeipel1910}
\bibinfo{author}{{von Zeipel}, H.}
\newblock \bibinfo{title}{{Sur l'application des s{\'e}ries de M. Lindstedt {\`a} l'{\'e}tude du mouvement des com{\`e}tes p{\'e}riodiques}}.
\newblock \emph{\bibinfo{journal}{Astronomische Nachrichten}} {\bibinfo{volume}{183}}, \bibinfo{pages}{345} (\bibinfo{year}{1910}).

\bibitem{Lidov1962}
\bibinfo{author}{{Lidov}, M.~L.}
\newblock \bibinfo{title}{{The evolution of orbits of artificial satellites of planets under the action of gravitational perturbations of external bodies}}.
\newblock \emph{\bibinfo{journal}{Planet. Space Sci.}} {\bibinfo{volume}{9}}, \bibinfo{pages}{719--759} (\bibinfo{year}{1962}).

\bibitem{Kozai1962}
\bibinfo{author}{{Kozai}, Y.}
\newblock \bibinfo{title}{{Secular perturbations of asteroids with high inclination and eccentricity}}.
\newblock \emph{\bibinfo{journal}{Astron. J.}} {\bibinfo{volume}{67}}, \bibinfo{pages}{591--598} (\bibinfo{year}{1962}).

\bibitem{2006ApJ...645.1152H}
\bibinfo{author}{{Hopman}, C.} \& \bibinfo{author}{{Alexander}, T.}
\newblock \bibinfo{title}{{Resonant Relaxation near a Massive Black Hole: The Stellar Distribution and Gravitational Wave Sources}}.
\newblock \emph{\bibinfo{journal}{Astrophys. J.}} {\bibinfo{volume}{645}}, \bibinfo{pages}{1152--1163} (\bibinfo{year}{2006}).

\bibitem{2018A&A...609A..26G}
\bibinfo{author}{{Gallego-Cano}, E.} \emph{et~al.}
\newblock \bibinfo{title}{{The distribution of stars around the Milky Way's central black hole. I. Deep star counts}}.
\newblock \emph{\bibinfo{journal}{Astron. Astrophys.}} {\bibinfo{volume}{609}}, \bibinfo{pages}{A26} (\bibinfo{year}{2018}).

\bibitem{Stephan2016}
\bibinfo{author}{{Stephan}, A.~P.} \emph{et~al.}
\newblock \bibinfo{title}{{Merging binaries in the Galactic Center: the eccentric Kozai-Lidov mechanism with stellar evolution}}.
\newblock \emph{\bibinfo{journal}{Mon. Not. R. Soc.}} {\bibinfo{volume}{460}}, \bibinfo{pages}{3494--3504} (\bibinfo{year}{2016}).

\bibitem{2003ApJ...582L.105S}
\bibinfo{author}{{Soker}, N.} \& \bibinfo{author}{{Tylenda}, R.}
\newblock \bibinfo{title}{{Main-Sequence Stellar Eruption Model for V838 Monocerotis}}.
\newblock \emph{\bibinfo{journal}{Astrophys. J.l}} {\bibinfo{volume}{582}}, \bibinfo{pages}{L105--L108} (\bibinfo{year}{2003}).

\bibitem{Alexander2014}
\bibinfo{author}{{Alexander}, T.} \& \bibinfo{author}{{Pfuhl}, O.}
\newblock \bibinfo{title}{{Constraining the Dark Cusp in the Galactic Center by Long-period Binaries}}.
\newblock \emph{\bibinfo{journal}{Astrophys. J.}} {\bibinfo{volume}{780}}, \bibinfo{pages}{148} (\bibinfo{year}{2014}).

\bibitem{Rose2020}
\bibinfo{author}{{Rose}, S.~C.} \emph{et~al.}
\newblock \bibinfo{title}{{On Socially Distant Neighbors: Using Binaries to Constrain the Density of Objects in the Galactic Center}}.
\newblock \emph{\bibinfo{journal}{Astrophys. J.}} {\bibinfo{volume}{904}}, \bibinfo{pages}{113} (\bibinfo{year}{2020}).

\bibitem{Ciurlo2023}
\bibinfo{author}{{Ciurlo}, A.} \emph{et~al.}
\newblock \bibinfo{title}{{The Swansong of the Galactic Center Source X7: An Extreme Example of Tidal Evolution near the Supermassive Black Hole}}.
\newblock \emph{\bibinfo{journal}{Astrophys. J.}} {\bibinfo{volume}{944}}, \bibinfo{pages}{136} (\bibinfo{year}{2023}).

\bibitem{stephan2019}
\bibinfo{author}{{Stephan}, A.~P.} \emph{et~al.}
\newblock \bibinfo{title}{{The Fate of Binaries in the Galactic Center: The Mundane and the Exotic}}.
\newblock \emph{\bibinfo{journal}{Astrophys. J.}} {\bibinfo{volume}{878}}, \bibinfo{pages}{58} (\bibinfo{year}{2019}).

\bibitem{Paxton2011}
\bibinfo{author}{{Paxton}, B.} \emph{et~al.}
\newblock \bibinfo{title}{{Modules for Experiments in Stellar Astrophysics (MESA)}}.
\newblock \emph{\bibinfo{journal}{Astrophys. J.s}} {\bibinfo{volume}{192}}, \bibinfo{pages}{3} (\bibinfo{year}{2011}).

\bibitem{Ott1999}
\bibinfo{author}{{Ott}, T.}, \bibinfo{author}{{Eckart}, A.} \& \bibinfo{author}{{Genzel}, R.}
\newblock \bibinfo{title}{{Variable and Embedded Stars in the Galactic Center}}.
\newblock \emph{\bibinfo{journal}{Astrophys. J.}} {\bibinfo{volume}{523}}, \bibinfo{pages}{248--264} (\bibinfo{year}{1999}).

\bibitem{Depoy2004}
\bibinfo{author}{{DePoy}, D.~L.} \emph{et~al.}
\newblock \bibinfo{title}{{The Nature of the Variable Galactic Center Source IRS 16SW}}.
\newblock \emph{\bibinfo{journal}{Astrophys. J.}} {\bibinfo{volume}{617}}, \bibinfo{pages}{1127--1130} (\bibinfo{year}{2004}).

\bibitem{Martins2006}
\bibinfo{author}{{Martins}, F.} \emph{et~al.}
\newblock \bibinfo{title}{{GCIRS 16SW: A Massive Eclipsing Binary in the Galactic Center}}.
\newblock \emph{\bibinfo{journal}{Astrophys. J.l}} {\bibinfo{volume}{649}}, \bibinfo{pages}{L103--L106} (\bibinfo{year}{2006}).

\bibitem{Paxton2019}
\bibinfo{author}{{Paxton}, B.} \emph{et~al.}
\newblock \bibinfo{title}{{Modules for Experiments in Stellar Astrophysics (MESA): Pulsating Variable Stars, Rotation, Convective Boundaries, and Energy Conservation}}.
\newblock \emph{\bibinfo{journal}{Astrophys. J.s}} {\bibinfo{volume}{243}}, \bibinfo{pages}{10} (\bibinfo{year}{2019}).

\bibitem{Vioque2018}
\bibinfo{author}{{Vioque}, M.}, \bibinfo{author}{{Oudmaijer}, R.~D.}, \bibinfo{author}{{Baines}, D.}, \bibinfo{author}{{Mendigut{\'\i}a}, I.} \& \bibinfo{author}{{P{\'e}rez-Mart{\'\i}nez}, R.}
\newblock \bibinfo{title}{{Gaia DR2 study of Herbig Ae/Be stars}}.
\newblock \emph{\bibinfo{journal}{Astron. Astrophys.}} {\bibinfo{volume}{620}}, \bibinfo{pages}{A128} (\bibinfo{year}{2018}).

\bibitem{peissker2023b}
\bibinfo{author}{{Pei{\ss}ker}, F.} \emph{et~al.}
\newblock \bibinfo{title}{{X3: A High-mass Young Stellar Object Close to the Supermassive Black Hole Sgr A*}}.
\newblock \emph{\bibinfo{journal}{Astrophys. J.}} {\bibinfo{volume}{944}}, \bibinfo{pages}{231} (\bibinfo{year}{2023b}).

\bibitem{peissker2021}
\bibinfo{author}{{Pei{\ss}ker}, F.} \emph{et~al.}
\newblock \bibinfo{title}{{First Observed Interaction of the Circumstellar Envelope of an S-star with the Environment of Sgr A*}}.
\newblock \emph{\bibinfo{journal}{Astrophys. J.}} {\bibinfo{volume}{909}}, \bibinfo{pages}{62} (\bibinfo{year}{2021a}).

\bibitem{Shahzamanian2016}
\bibinfo{author}{{Shahzamanian}, B.} \emph{et~al.}
\newblock \bibinfo{title}{{Polarized near-infrared light of the Dusty S-cluster Object (DSO/G2) at the Galactic center}}.
\newblock \emph{\bibinfo{journal}{Astron. Astrophys.}} {\bibinfo{volume}{593}}, \bibinfo{pages}{A131} (\bibinfo{year}{2016}).

\bibitem{2017ApJ...847..120H}
\bibinfo{author}{{Habibi}, M.} \emph{et~al.}
\newblock \bibinfo{title}{{Twelve Years of Spectroscopic Monitoring in the Galactic Center: The Closest Look at S-stars near the Black Hole}}.
\newblock \emph{\bibinfo{journal}{Astrophys. J.}} {\bibinfo{volume}{847}}, \bibinfo{pages}{120} (\bibinfo{year}{2017}).

\bibitem{Najarro1997}
\bibinfo{author}{{Najarro}, F.} \emph{et~al.}
\newblock \bibinfo{title}{{Quantitative spectroscopy of the HeI cluster in the Galactic center.}}
\newblock \emph{\bibinfo{journal}{Astron. Astrophys.}} {\bibinfo{volume}{325}}, \bibinfo{pages}{700--708} (\bibinfo{year}{1997}).

\bibitem{McKee2007}
\bibinfo{author}{{McKee}, C.~F.} \& \bibinfo{author}{{Ostriker}, E.~C.}
\newblock \bibinfo{title}{{Theory of Star Formation}}.
\newblock \emph{\bibinfo{journal}{Annu. Rev. Astron. Astrophys.}} {\bibinfo{volume}{45}}, \bibinfo{pages}{565--687} (\bibinfo{year}{2007}).

\bibitem{Wichittanakom2020}
\bibinfo{author}{{Wichittanakom}, C.} \emph{et~al.}
\newblock \bibinfo{title}{{The accretion rates and mechanisms of Herbig Ae/Be stars}}.
\newblock \emph{\bibinfo{journal}{Mon. Not. R. Soc.}} {\bibinfo{volume}{493}}, \bibinfo{pages}{234--249} (\bibinfo{year}{2020}).

\bibitem{Gaia2016}
\bibinfo{author}{{Gaia Collaboration}} \emph{et~al.}
\newblock \bibinfo{title}{{The Gaia mission}}.
\newblock \emph{\bibinfo{journal}{Astron. Astrophys.}} {\bibinfo{volume}{595}}, \bibinfo{pages}{A1} (\bibinfo{year}{2016}).

\bibitem{Gaia2018}
\bibinfo{author}{{Gaia Collaboration}} \emph{et~al.}
\newblock \bibinfo{title}{{Gaia Data Release 2. Summary of the contents and survey properties}}.
\newblock \emph{\bibinfo{journal}{Astron. Astrophys.}} {\bibinfo{volume}{616}}, \bibinfo{pages}{A1} (\bibinfo{year}{2018}).

\bibitem{Schoedel2002}
\bibinfo{author}{{Sch{\"o}del}, R.} \emph{et~al.}
\newblock \bibinfo{title}{{A star in a 15.2-year orbit around the supermassive black hole at the centre of the Milky Way}}.
\newblock \emph{\bibinfo{journal}{Nature}} {\bibinfo{volume}{419}}, \bibinfo{pages}{694--696} (\bibinfo{year}{2002}).

\bibitem{Do2019S2}
\bibinfo{author}{{Do}, T.} \emph{et~al.}
\newblock \bibinfo{title}{{Relativistic redshift of the star S0-2 orbiting the Galactic Center supermassive black hole}}.
\newblock \emph{\bibinfo{journal}{Science}} {\bibinfo{volume}{365}}, \bibinfo{pages}{664--668} (\bibinfo{year}{2019}).

\bibitem{Parsa2017}
\bibinfo{author}{{Parsa}, M.} \emph{et~al.}
\newblock \bibinfo{title}{{Investigating the Relativistic Motion of the Stars Near the Supermassive Black Hole in the Galactic Center}}.
\newblock \emph{\bibinfo{journal}{Astrophys. J.}} {\bibinfo{volume}{845}}, \bibinfo{pages}{22} (\bibinfo{year}{2017}).

\bibitem{Liu1989}
\bibinfo{author}{Liu, D.~C.} \& \bibinfo{author}{Nocedal, J.}
\newblock \bibinfo{title}{On the limited memory bfgs method for large scale optimization}.
\newblock \emph{\bibinfo{journal}{Mathematical Programming}} {\bibinfo{volume}{45}}, \bibinfo{pages}{503--528} (\bibinfo{year}{1989}).
\newblock \urlprefix\url{https://api.semanticscholar.org/CorpusID:5681609}.

\bibitem{Zhu1997}
\bibinfo{author}{Zhu, C.}, \bibinfo{author}{Byrd, R.~H.}, \bibinfo{author}{Lu, P.} \& \bibinfo{author}{Nocedal, J.}
\newblock \bibinfo{title}{Algorithm 778: L-bfgs-b: Fortran subroutines for large-scale bound-constrained optimization.}
\newblock \emph{\bibinfo{journal}{ACM Trans. Math. Softw.}} {\bibinfo{volume}{23}}, \bibinfo{pages}{550--560} (\bibinfo{year}{1997}).
\newblock \urlprefix\url{http://dblp.uni-trier.de/db/journals/toms/toms23.html#ZhuBLN97}.

\bibitem{Foreman-Mackey2013}
\bibinfo{author}{{Foreman-Mackey}, D.}, \bibinfo{author}{{Hogg}, D.~W.}, \bibinfo{author}{{Lang}, D.} \& \bibinfo{author}{{Goodman}, J.}
\newblock \bibinfo{title}{{emcee: The MCMC Hammer}}.
\newblock \emph{\bibinfo{journal}{Publ. Astron. Soc. Pac.}} {\bibinfo{volume}{125}}, \bibinfo{pages}{306} (\bibinfo{year}{2013}).

\bibitem{Sabha2012}
\bibinfo{author}{{Sabha}, N.} \emph{et~al.}
\newblock \bibinfo{title}{{The S-star cluster at the center of the Milky Way. On the nature of diffuse NIR emission in the inner tenth of a parsec}}.
\newblock \emph{\bibinfo{journal}{Astron. Astrophys.}} {\bibinfo{volume}{545}}, \bibinfo{pages}{A70} (\bibinfo{year}{2012}).

\bibitem{Peissker2020c}
\bibinfo{author}{{Pei{\ss}ker}, F.}, \bibinfo{author}{{Eckart}, A.}, \bibinfo{author}{{Sabha}, N.~B.}, \bibinfo{author}{{Zaja{\v{c}}ek}, M.} \& \bibinfo{author}{{Bhat}, H.}
\newblock \bibinfo{title}{{Near- and Mid-infrared Observations in the Inner Tenth of a Parsec of the Galactic Center Detection of Proper Motion of a Filament Very Close to Sgr A*}}.
\newblock \emph{\bibinfo{journal}{Astrophys. J.}} {\bibinfo{volume}{897}}, \bibinfo{pages}{28} (\bibinfo{year}{2020c}).

\bibitem{Ciurlo2021}
\bibinfo{author}{{Ciurlo}, A.} \emph{et~al.}
\newblock \bibinfo{title}{{Upper Limit on Brackett-{\ensuremath{\gamma}} Emission from the Immediate Accretion Flow onto the Galactic Black Hole}}.
\newblock \emph{\bibinfo{journal}{Astrophys. J.}} {\bibinfo{volume}{910}}, \bibinfo{pages}{143} (\bibinfo{year}{2021}).

\bibitem{Peissker2019}
\bibinfo{author}{{Pei{\ss}ker}, F.} \emph{et~al.}
\newblock \bibinfo{title}{{New bow-shock source with bipolar morphology in the vicinity of Sgr A*}}.
\newblock \emph{\bibinfo{journal}{Astron. Astrophys.}} {\bibinfo{volume}{624}}, \bibinfo{pages}{A97} (\bibinfo{year}{2019}).

\bibitem{Freudling2013}
\bibinfo{author}{{Freudling}, W.} \emph{et~al.}
\newblock \bibinfo{title}{{Automated data reduction workflows for astronomy. The ESO Reflex environment}}.
\newblock \emph{\bibinfo{journal}{Astron. Astrophys.}} {\bibinfo{volume}{559}}, \bibinfo{pages}{A96} (\bibinfo{year}{2013}).

\end{thebibliography}

\begin{thebibliography}{10}
\expandafter\ifx\csname url\endcsname\relax
  \def\url#1{\burl{#1}}\fi
\expandafter\ifx\csname urlprefix\endcsname\relax\def\urlprefix{URL }\fi
\providecommand{\bibinfo}[2]{#2}
\providecommand{\eprint}[2][]{\url{#2}}
\providecommand{\doi}[1]{\url{https://doi.org/#1}}
\bibcommenthead

\bibitem{Schoedel2002}
\bibinfo{author}{{Sch{\"o}del}, R.} \emph{et~al.}
\newblock \bibinfo{title}{{A star in a 15.2-year orbit around the supermassive black hole at the centre of the Milky Way}}.
\newblock \emph{\bibinfo{journal}{Nature}} \textbf{\bibinfo{volume}{419}}, \bibinfo{pages}{694--696} (\bibinfo{year}{2002}).

\bibitem{Gillessen2009}
\bibinfo{author}{{Gillessen}, S.} \emph{et~al.}
\newblock \bibinfo{title}{{Monitoring Stellar Orbits Around the Massive Black Hole in the Galactic Center}}.
\newblock \emph{\bibinfo{journal}{Astrophys. J.}} \textbf{\bibinfo{volume}{692}}, \bibinfo{pages}{1075--1109} (\bibinfo{year}{2009}).

\bibitem{peissker2023c}
\bibinfo{author}{{Pei{\ss}ker}, F.} \emph{et~al.}
\newblock \bibinfo{title}{{The Evaporating Massive Embedded Stellar Cluster IRS 13 Close to Sgr A*. I. Detection of a Rich Population of Dusty Objects in the IRS 13 Cluster}}.
\newblock \emph{\bibinfo{journal}{Astrophys. J.}} \textbf{\bibinfo{volume}{956}}, \bibinfo{pages}{70} (\bibinfo{year}{2023c}).

\bibitem{Eisenhauer2005}
\bibinfo{author}{{Eisenhauer}, F.} \emph{et~al.}
\newblock \bibinfo{title}{{SINFONI in the Galactic Center: Young Stars and Infrared Flares in the Central Light-Month}}.
\newblock \emph{\bibinfo{journal}{Astrophys. J.}} \textbf{\bibinfo{volume}{628}}, \bibinfo{pages}{246--259} (\bibinfo{year}{2005}).

\bibitem{Ali2020}
\bibinfo{author}{{Ali}, B.} \emph{et~al.}
\newblock \bibinfo{title}{{Kinematic Structure of the Galactic Center S Cluster}}.
\newblock \emph{\bibinfo{journal}{Astrophys. J.}} \textbf{\bibinfo{volume}{896}}, \bibinfo{pages}{100} (\bibinfo{year}{2020}).

\bibitem{Ott1999}
\bibinfo{author}{{Ott}, T.}, \bibinfo{author}{{Eckart}, A.} \& \bibinfo{author}{{Genzel}, R.}
\newblock \bibinfo{title}{{Variable and Embedded Stars in the Galactic Center}}.
\newblock \emph{\bibinfo{journal}{Astrophys. J.}} \textbf{\bibinfo{volume}{523}}, \bibinfo{pages}{248--264} (\bibinfo{year}{1999}).

\bibitem{Viehmann2006}
\bibinfo{author}{{Viehmann}, T.}, \bibinfo{author}{{Eckart}, A.}, \bibinfo{author}{{Sch{\"o}del}, R.}, \bibinfo{author}{{Pott}, J.~U.} \& \bibinfo{author}{{Moultaka}, J.}
\newblock \bibinfo{title}{{Dusty Sources at the Galactic Center the N- and Q-Band Views with VISIR}}.
\newblock \emph{\bibinfo{journal}{Astrophys. J.}} \textbf{\bibinfo{volume}{642}}, \bibinfo{pages}{861--867} (\bibinfo{year}{2006}).

\bibitem{Gautam2024}
\bibinfo{author}{{Gautam}, A.~K.} \emph{et~al.}
\newblock \bibinfo{title}{{An Estimate of the Binary Star Fraction Among Young Stars at the Galactic Center: Possible Evidence of a Radial Dependence}}.
\newblock \emph{\bibinfo{journal}{arXiv e-prints}} \bibinfo{pages}{arXiv:2401.12555} (\bibinfo{year}{2024}).

\bibitem{Peissker2024}
\bibinfo{author}{{Pei{\ss}ker}, F.} \emph{et~al.}
\newblock \bibinfo{title}{{Candidate young stellar objects in the S-cluster: Kinematic analysis of a subpopulation of the low-mass G objects close to Sgr A*}}.
\newblock \emph{\bibinfo{journal}{Astron. Astrophys.}} \textbf{\bibinfo{volume}{686}}, \bibinfo{pages}{A235} (\bibinfo{year}{2024}).

\bibitem{Labadie2011}
\bibinfo{author}{{Labadie}, L.} \emph{et~al.}
\newblock \bibinfo{title}{{High-contrast optical imaging of companions: the case of the brown dwarf binary HD 130948 BC}}.
\newblock \emph{\bibinfo{journal}{Astron. Astrophys.}} \textbf{\bibinfo{volume}{526}}, \bibinfo{pages}{A144} (\bibinfo{year}{2011}).

\bibitem{Habibi2017}
\bibinfo{author}{{Habibi}, M.} \emph{et~al.}
\newblock \bibinfo{title}{{Twelve Years of Spectroscopic Monitoring in the Galactic Center: The Closest Look at S-stars near the Black Hole}}.
\newblock \emph{\bibinfo{journal}{Astrophys. J.}} \textbf{\bibinfo{volume}{847}}, \bibinfo{pages}{120} (\bibinfo{year}{2017}).

\bibitem{Schoedel2010}
\bibinfo{author}{{Sch{\"o}del}, R.}, \bibinfo{author}{{Najarro}, F.}, \bibinfo{author}{{Muzic}, K.} \& \bibinfo{author}{{Eckart}, A.}
\newblock \bibinfo{title}{{Peering through the veil: near-infrared photometry and extinction for the Galactic nuclear star cluster. Accurate near infrared H, Ks, and L' photometry and the near-infrared extinction-law toward the central parsec of the Galaxy}}.
\newblock \emph{\bibinfo{journal}{Astron. Astrophys.}} \textbf{\bibinfo{volume}{511}}, \bibinfo{pages}{A18} (\bibinfo{year}{2010}).

\bibitem{Sabha2012}
\bibinfo{author}{{Sabha}, N.} \emph{et~al.}
\newblock \bibinfo{title}{{The S-star cluster at the center of the Milky Way. On the nature of diffuse NIR emission in the inner tenth of a parsec}}.
\newblock \emph{\bibinfo{journal}{Astron. Astrophys.}} \textbf{\bibinfo{volume}{545}}, \bibinfo{pages}{A70} (\bibinfo{year}{2012}).

\bibitem{Fritz2011}
\bibinfo{author}{{Fritz}, T.~K.} \emph{et~al.}
\newblock \bibinfo{title}{{Line Derived Infrared Extinction toward the Galactic Center}}.
\newblock \emph{\bibinfo{journal}{ApJ}} \textbf{\bibinfo{volume}{737}}, \bibinfo{pages}{73} (\bibinfo{year}{2011}).

\bibitem{peissker2023b}
\bibinfo{author}{{Pei{\ss}ker}, F.} \emph{et~al.}
\newblock \bibinfo{title}{{X3: A High-mass Young Stellar Object Close to the Supermassive Black Hole Sgr A*}}.
\newblock \emph{\bibinfo{journal}{Astrophys. J.}} \textbf{\bibinfo{volume}{944}}, \bibinfo{pages}{231} (\bibinfo{year}{2023b}).

\bibitem{Whitney2004}
\bibinfo{author}{{Whitney}, B.~A.}, \bibinfo{author}{{Indebetouw}, R.}, \bibinfo{author}{{Bjorkman}, J.~E.} \& \bibinfo{author}{{Wood}, K.}
\newblock \bibinfo{title}{{Two-Dimensional Radiative Transfer in Protostellar Envelopes. III. Effects of Stellar Temperature}}.
\newblock \emph{\bibinfo{journal}{Astrophys. J.}} \textbf{\bibinfo{volume}{617}}, \bibinfo{pages}{1177--1190} (\bibinfo{year}{2004}).

\bibitem{Keppler2018}
\bibinfo{author}{{Keppler}, M.} \emph{et~al.}
\newblock \bibinfo{title}{{Discovery of a planetary-mass companion within the gap of the transition disk around PDS 70}}.
\newblock \emph{\bibinfo{journal}{Astron. Astrophys.}} \textbf{\bibinfo{volume}{617}}, \bibinfo{pages}{A44} (\bibinfo{year}{2018}).

\bibitem{Garcia2013}
\bibinfo{author}{{Garcia}, P.~J.~V.} \emph{et~al.}
\newblock \bibinfo{title}{{Pre-main-sequence binaries with tidally disrupted discs: the Br{\ensuremath{\gamma}} in HD 104237}}.
\newblock \emph{\bibinfo{journal}{Mon. Not. R. Soc.}} \textbf{\bibinfo{volume}{430}}, \bibinfo{pages}{1839--1853} (\bibinfo{year}{2013}).

\bibitem{Dinh2024}
\bibinfo{author}{{Dinh}, C.~K.} \emph{et~al.}
\newblock \bibinfo{title}{{High-resolution, Mid-infrared Color Temperature Mapping of the Central 10$^{''}$ of the Galaxy}}.
\newblock \emph{\bibinfo{journal}{Astron. J.}} \textbf{\bibinfo{volume}{167}}, \bibinfo{pages}{41} (\bibinfo{year}{2024}).

\bibitem{Shimizu2023}
\bibinfo{author}{{Shimizu}, T.}, \bibinfo{author}{{Uyama}, T.}, \bibinfo{author}{{Hori}, Y.}, \bibinfo{author}{{Tamura}, M.} \& \bibinfo{author}{{Wallack}, N.}
\newblock \bibinfo{title}{{High-contrast Imaging around a 2 Myr-old CI Tau with a Close-in Gas Giant}}.
\newblock \emph{\bibinfo{journal}{Astron. J.}} \textbf{\bibinfo{volume}{165}}, \bibinfo{pages}{20} (\bibinfo{year}{2023}).

\bibitem{Stapper2022}
\bibinfo{author}{{Stapper}, L.~M.}, \bibinfo{author}{{Hogerheijde}, M.~R.}, \bibinfo{author}{{van Dishoeck}, E.~F.} \& \bibinfo{author}{{Mentel}, R.}
\newblock \bibinfo{title}{{The mass and size of Herbig disks as seen by ALMA}}.
\newblock \emph{\bibinfo{journal}{Astron. Astrophys.}} \textbf{\bibinfo{volume}{658}}, \bibinfo{pages}{A112} (\bibinfo{year}{2022}).

\bibitem{Peissker2020c}
\bibinfo{author}{{Pei{\ss}ker}, F.}, \bibinfo{author}{{Eckart}, A.}, \bibinfo{author}{{Sabha}, N.~B.}, \bibinfo{author}{{Zaja{\v{c}}ek}, M.} \& \bibinfo{author}{{Bhat}, H.}
\newblock \bibinfo{title}{{Near- and Mid-infrared Observations in the Inner Tenth of a Parsec of the Galactic Center Detection of Proper Motion of a Filament Very Close to Sgr A*}}.
\newblock \emph{\bibinfo{journal}{Astrophys. J.}} \textbf{\bibinfo{volume}{897}}, \bibinfo{pages}{28} (\bibinfo{year}{2020c}).

\bibitem{gravity2018}
\bibinfo{author}{{Gravity Collaboration}} \emph{et~al.}
\newblock \bibinfo{title}{{Detection of the gravitational redshift in the orbit of the star S2 near the Galactic centre massive black hole}}.
\newblock \emph{\bibinfo{journal}{Astron. Astrophys.}} \textbf{\bibinfo{volume}{615}}, \bibinfo{pages}{L15} (\bibinfo{year}{2018}).

\bibitem{Newberry1991}
\bibinfo{author}{{Newberry}, M.~V.}
\newblock \bibinfo{title}{{Signal-to-Noise Considerations for Sky-Subtracted CCD Data}}.
\newblock \emph{\bibinfo{journal}{Publ. Astron. Soc. Pacif.}} \textbf{\bibinfo{volume}{103}}, \bibinfo{pages}{122} (\bibinfo{year}{1991}).

\bibitem{Gautam2019}
\bibinfo{author}{{Gautam}, A.~K.} \emph{et~al.}
\newblock \bibinfo{title}{{An Adaptive Optics Survey of Stellar Variability at the Galactic Center}}.
\newblock \emph{\bibinfo{journal}{Astrophys. J.}} \textbf{\bibinfo{volume}{871}}, \bibinfo{pages}{103} (\bibinfo{year}{2019}).

\bibitem{Pfuhl2014}
\bibinfo{author}{{Pfuhl}, O.} \emph{et~al.}
\newblock \bibinfo{title}{{Massive Binaries in the Vicinity of Sgr A*}}.
\newblock \emph{\bibinfo{journal}{Astrophys. J.}} \textbf{\bibinfo{volume}{782}}, \bibinfo{pages}{101} (\bibinfo{year}{2014}).

\end{thebibliography}

\end{document}